\documentclass[aps,prd,twocolumn,showpacs,superscriptaddress,groupedaddress,nofootinbib]{revtex4-2}  

\usepackage{graphicx}
\usepackage{subfigure}
\usepackage{amsmath} 
\usepackage[colorlinks=true,citecolor=blue,urlcolor=magenta,breaklinks]{hyperref}
\usepackage{times}
\usepackage[utf8]{inputenc}
\usepackage[english]{babel}
\usepackage{natbib}

\graphicspath{{./}{figures/}}
\usepackage[dvipsnames]{xcolor}

\def\araa{ARA\&A}             
\def\apj{ApJ}                 
\def\apjs{ApJS}              
\def\aap{A\&A}                
\def\jcap{JCAP}              %
\def\mnras{MNRAS}             
\def\prc{Phys.~Rev.~C}        
\def\prd{Phys.~Rev.~D}        
\def\prl{Phys.~Rev.~Lett.}    
\def\nat{Nature}              

\begin{document}
\title{Linking solar bosonic dark matter halos  and active neutrinos}
\author{Il\'idio Lopes}
\email[]{ilidio.lopes@tecnico.ulisboa.pt}
\affiliation{ Centro de Astrof\'{\i}sica e Gravita\c c\~ao - CENTRA,
	Departamento de F\'{\i}sica, Instituto Superior T\'ecnico - IST,\\ 
	Universidade de Lisboa - UL, Avenida Rovisco Pais 1, 1049-001 Lisboa, Portugal \\}
 
\begin{abstract}
Our study investigates the complex interaction between active neutrinos and the ultralight bosonic dark matter halo surrounding the Sun. This halo extends over several solar radii due to the Sun's gravitational field, and we represent it as a coherent oscillating classical field configuration of bosonic dark matter particles that vary in time. 
Our investigation has revealed that, based on the available solar neutrino flux data, these novel models do not surpass the performance of the conventional neutrino flavour oscillation model. Furthermore, we discuss how next-generation solar neutrino detectors have the potential to provide evidence for the existence or absence of the ultralight dark matter halo.
\end{abstract}


\maketitle

\section{Introduction} \label{sec:intro}
While studying the movement of galaxies in the Coma cluster, Fritz Zwicky became the first astronomer to detect a discrepancy between visible matter and gravitational forces. In a groundbreaking 1933 article, he presented a compelling finding: the visible matter's total mass in the cluster was insufficient to gravitationally bind the galaxies, identifying what is now known as the dark matter problem \citep{1933AcHPh...6..110Z}. 
Since then, many solutions have continuously been put forward to solve this problem, including invisible massive astronomical bodies such as black holes, alternative theories to general gravity, extensions to the standard model of fundamental particles and interactions postulating the existence of new particles. Although many articles are being published on these very active research fields, the latter class of solutions is the most popular and widely accepted. However, identifying the definitive new particle or particles remains a significant challenge
\citep{2017arXiv170704591B}.  
So far, the theoretical and experimental efforts to discover such particles have focused primarily on the so-called weakly interacting massive particles (WIMPs). Accordingly, these particles are neutral and nonrelativistic particles, with masses varying from a few  GeV to    $10^3$ GeV and interacting ultraweakly with the standard particles, e.g., \citep{2010ARA&A..48..495F,2015PhR...555....1B}.

\medskip\noindent
The lack of evidence for the existence of WIMPs and the need to answer unsettled problems in the current standard cosmological model —
Lambda cold dark matter model — motivated the study of properties of ultralight particles as viable dark matter candidates. These particles are well motivated by modern theories (e.g.,\citep{2021EPJC...81.1015A}), many of which predict the existence of spin-0 and spin-1 bosons, including axions or axionlike particles and dark photons \citep{2021EPJC...81.1015A,2015ARNPS..65..485G,2009PhRvD..79b3519A}.   These ultralight particles collectively behave like a bosonic field. In this work, we call all these particles, including the classical axion \citep{2020PhR...870....1D} or their closest relative,  the axionlike particles \citep{2016PhR...643....1M}, simply  "bosons" or "ultra-light particles" if not stated otherwise. Interestingly, we should be able to detect such fields by their interaction with standard particles in the future.
A cornucopia of experiments aims to detect such bosonic particles by the emission of photons created by their interaction with magnetic fields \citep{2010PhRvL.104d1301A}, nuclear magnetic resonance (e.g.,\citep{2014PhRvX...4b1030B})  and axion spin precession
(e.g.,\citep{2014PhRvD..89d3522S}).   

\medskip\noindent   
Here, we discuss the possibility of this ultralight dark matter particle interacting with solar neutrinos using a well-established model for predicting neutrino flavor oscillation through vacuum and matter. In this work, we discuss also how current and future solar neutrino detectors (e.g.,\citep{2021ARNPS..71..491O})  could be used to constrain the properties of these  particles.  

\medskip\noindent
The article is organized as follows: Sec. \ref{sec-Axion}  presents the properties of the local dark matter field. Section \ref{sec-ALDM} explains the mechanism by which the local dark matter field interacts with the active neutrinos; Section \ref{sec-PLDM}  introduces the survival probability of electron neutrino function in the standard three-neutrino oscillation flavour model; Section  \ref{sec-PLDM2} generalizes the result of the previous section to the new neutrino dark matter model.  Section  \ref{sec-LDMeneutrinos}  discusses the time dependence of the survival probability of electron neutrinos. Section  \ref{sec-LDMeneutrinos2} gives predictions of the new neutrino model for the present Sun; section  \ref{sec-Summary}  summarises the main results and conclusions of this work.

\medskip\noindent   
If not stated otherwise, we work in natural units ($c=\hbar=1$). All standard units are expressed in GeV by applying the usual conversion rules.  These are the most common ones used in this work: $\rm 1\, m = 5.068 \times 10^{15} GeV^{-1}$, $\rm 1\, kg = 5.610 \times 10^{26} GeV$ and $\rm 1\, sec = 1.519 \times 10^{24} GeV^{-1}$.

\section{ultra-light dark matter}
\label{sec-Axion}

\subsection{The origin of the ultralight  dark matter field}  
\medskip\noindent
If boson particles exist today,  they were produced abundantly in the early Universe. The production of light-dark matter can take many forms. We have thermal production, dark matter decay, parametric resonance, and topological defect decay, among other mechanisms (e.g.,\citep{2021PhRvD.103k5004D,2022RPPh...85e6201B}). For the light-dark matter field, some authors obtained  $ \Omega_\psi^2 =0.1 \left({a_o}/{10^{17}\; {\rm GeV}}\right)^2
\left({m_\psi}/{10^{-22}\; {\rm eV}}\right)^{1/2}$, where $a_o$ is a parameter that relates to the initial misalignment angle of the axion, and $m_\psi$  is the axion mass \citep{2017PhRvD..95d3541H,2020PrPNP.11303787N}.
Therefore, we will assume that at least a fraction of the dark matter {\it background} in the solar neighborhood  is made of an ensemble of ultralight bosons:
\begin{eqnarray}	
	\bar{\rho}_{b\psi}= \bar{\rho}^\odot_{\rm DM}\left(\frac{\Omega_{\psi} h^2}{\Omega_{\rm DM}h^2}\right)
\end{eqnarray}
where  $\bar{\rho}^\odot_{\rm DM}$ is the local density of dark matter in the solar neighborhood, $\Omega_{\rm DM}^2 h^2 $ and $\Omega_\psi^2 h^2 $ are the total DM and axion energy density parameters in the present Universe, and $h$ is the reduced Hubble constant such $h\equiv H_0/({\rm 100\, km\, s^{-1}\, Mpc^{-1}})$.
Recent measurements of the dark matter constituents give  
$\rho^\odot_{\rm DM}=0.39\; GeV \; cm^{-3} $ \citep{2010JCAP...08..004C} and  $\Omega_{\rm DM}h^2=0.12$ \citep{2020A&A...641A...6P}.
Now, we compute  the  averaged density number of dark matter $\bar{n}_{b\psi}$  near the Sun  as the ratio between  the averaged local density  $\bar{\rho}_{b\psi}$ and $m_\psi$, such that
$\bar{n}_{b\psi}=\bar{\rho}_{b\psi}/m_\psi$. 
For example, considering axions with \( m_\psi = 10^{-22} \, \text{eV} \) and \( \Omega_{\psi} h^2 = \Omega_{\text{DM}} h^2 \), we obtain \( \bar{n}_{b\psi} = 3.9 \times 10^{30} \, \text{cm}^{-3} \).
This value is only two orders of magnitude smaller than the density of electrons in the Sun's core, $\bar{n}_e\sim 6\;10^{31}\;cm^{-3}$ \citep{2013ApJ...765...14L}. 

\medskip\noindent 
Ultralight boson particles behave as nonrelativistic matter and account for dark matter in the present Universe. Moreover, such a population of particles will smooth inhomogeneities in the dark matter distribution on scales smaller than the de Broglie wavelength $\lambda_{\rm dB}$.  
We have calculated the de Broglie wavelength, which is given by the equation $\lambda_{\rm dB}=1.24\times 10^{22} \left(10^{-23}{\rm eV}/m_\psi\right) \left(10^{-3}/v_\psi\right)\; {\rm cm} $, where $v_\psi$ is the virial velocity of the boson in the halo. For the fiducial boson, assuming $m_\psi=1.5\times 10^{-14}\;{\rm eV}$ and $v_\psi\sim 10^{-3}$, we obtain $\lambda_{\rm dB}=119 \;{\rm R_\odot}$.
Collectively, such particles form a dark matter background that we choose to represent as a coherent oscillating classical field configuration
(e.g.,\citep{1983PhLB..120..133A,2014JCAP...02..019K,2017PhRvL.118z1102B}).  Accordingly, the general form of this field reads
\begin{eqnarray}
	\psi_b (\vec{r},t)=\psi_{bo}\cos{\left(m_\psi t+\epsilon_o\right)}
	\label{eq:psi1} 
\end{eqnarray}
where $\psi_{bo}$ and $\epsilon_o$ are the amplitude and phase of this bosonic field background.  The amplitude  $\psi_{bo}$ is computed from the density $\bar{\rho}_{b\psi}$ as $\psi_{bo}=\sqrt{2\bar{\rho}_{b\psi}}/m_\psi$ and we assume that $\epsilon_o \approx 0$ \citep{2014JCAP...02..019K}.  In the determination of $\psi_b (\vec{r},t)$, since $v_\psi\sim 10^{-3}$ is very small (e.g.,\citep{2017PhRvL.118z1102B}), 
we neglect its  contribution for Eq. \eqref{eq:psi1}.

\medskip\noindent 
Here, we hypothesize that ultralight bosons around the Sun form a halo of dark matter.
As a consequence of such a gravitational bound system, we describe the boson field 
of this bound system similar to Eq. \eqref{eq:psi1}.
Conveniently, we define $\bar{\rho}_\psi$ as the average dark matter density inside the halo.
Since  $\bar{\rho}_\psi$  is larger than background density $\bar{\rho}_{b\psi}$,  we have $\bar{\rho}_\psi=\chi_\psi\bar{\rho}_{b\psi}$ where $\chi_\psi$  is a positive number larger than 1.  The boson halo, we are considering, shares similar properties to boson stars (e.g.,\citep{1986PhRvL..57.2485C,2012LRR....15....6L}), except that it is bounded by a gravitational potential of the host star, the Sun, rather than its self-gravity. Since these particles in the halo are maintained and stabilized by the gravitational potential of the host star, the total mass of halo $M_{\psi}$ must be much smaller than the mass of the star $M_{\odot}$, i.e., $M_{\psi} < M_{\odot}$, so specifically we can consider that $M_{\psi}\le M_{\odot}/2$.

\subsection{The local ultralight  dark matter field}  
\medskip\noindent 
Our study focuses on a halo of ultralight bosons hosted by the Sun, which we assume to be a spherical object with a total radius $R_{\psi}$ large enough to encompass the entire Sun. For convenience, we use a fiducial radius several times greater than the Sun's. In the nonrelativistic limit, we compute the boson field inside of the halo similar to  boson stars as originally proposed by
\citet{1968PhRv..172.1331K} and \citet{1969PhRv..187.1767R}
following in the footsteps of \citet{1955PhRv...97..511W}.
A recent review of the properties of boson stars can be found in \citet{2019RvMP...91d1002B}. 
Within the non-relativist effective field theory framework, \citet{2018PhRvD..98a6011N} derived an exact connection between the boson field, a real function, and a complex function known in this context as a  relaxion wave function.  \citet{2011PhRvD..84d3531C}   found that an exponential form is a suitable ansatz for describing the radial variation of the density profile inside the halo. \citet{2018PhRvD..98l3013E} has shown this form to be a good fit for the numerical solution to the
Gross-Pitäevskii-Poisson equation \citep{2007JCAP...06..025B}.  In our case, we consider that the boson field decays exponentially with the distance $r$ from the star's center  \citep{2011PhRvD..84d3531C}. Therefore,  the boson field created by the halo of dark matter particles in the presence of an external gravitational source   reads
\begin{eqnarray}
	\psi (r,t)=\sqrt{\frac{M_{\psi}}{\pi m_\psi R_{\psi}^3}}\;
	e^{-\frac{r}{R_{\psi}}}
	\cos{\left(m_\psi t\right)}.
	\label{eq:psi2} 
\end{eqnarray}
In this configuration, the radius of a  halo $R_{\psi}$ is determined by the gravitational potential of the external source: 
 $R_{\psi}=(M_{p}^2/M_\odot)m_\psi^{-2}$ where $M_{p}=(\hbar c/G)^{1/2}=1.2\times 10^{19}\;  GeV$ is the Planck mass, or  $R_{\psi}=3.7863\;(1\times 10^{-13} \;{\rm eV}/ m_\psi)^{2}\;R_\odot$.
Our study will focus on dark matter halos with a radius of at least 2 times that of the solar radius ($R_\psi \ge 2 R_\odot$), which implies the presence of dark matter particles with a mass $m_\psi$ lower than the critical threshold of $m^c_\psi \approx 3.6 \times 10^{-13} \mathrm{eV}$.
The boson field  $\psi (\vec{r},t)$ [Eq. \eqref{eq:psi2}] inside the halo behaves similar to the boson background field $\psi_{b} (\vec{r},t)$ [Eq. \eqref{eq:psi1}]. Both fields are oscillating in time with a frequency approximately equal to the boson mass $m_\psi$ (e.g.,\citep{2020JHEP...09..004B}).  We also found that  the density profile of the  dark matter halo (for $r\le R_{\psi}$ with $R_{\psi}\ge R_\odot$) reads
\begin{eqnarray}
	\rho_{\psi}(r)= \rho_{c\psi}\;
	e^{-\frac{2r}{R_{\psi}}},
\label{eq:rhopsi}	
\end{eqnarray} 
where $\rho_{c\psi}=M_\psi/(\pi R_\psi^3)$. 
This expression is identical to the one found by \citet{2020CmPhy...3....1B} for relaxion stars.  

\medskip\noindent 
We compute the  overdensity of particles inside the halo in a similar way  to the calculation done for the boson star \citep{2020CmPhy...3....1B}: the averaged density $\bar{\rho}_\psi$  of the halo is determined in comparison to the background density of the dark matter $\bar{\rho}_{b\phi}$, in such a way that the parameter of condensation $\chi_\psi$ corresponds to
\begin{eqnarray}
	\chi_\psi=\frac{\bar{\rho}_\psi}{\bar{\rho}_{b\psi}}
	\approx
	\frac{M_{\psi}}{ \pi R_{\psi}^3}\frac{1}{\bar{\rho}_{b\psi}}.
\end{eqnarray}
If we set $\bar{\rho}_{b\phi}=\bar{\rho}_{\rm DM}^{\odot}$ and express $M_\psi$ in units of solar masses, the previous equation can be rewritten as follows:
\begin{eqnarray}
	\chi_\psi = 4.987 \times 10^{22}
	\left(\frac{M_\psi}{M_\odot}\right)
	\left(\frac{m_\psi}{1\times 10^{-13}\;{\rm eV}}\right)^6,
\end{eqnarray}
where $m_\psi$ does not exceed the critical threshold value of $m^c_\psi$.

\medskip\noindent 
The standard solar model, which is based on solar neutrino fluxes and helioseismology data, provides an accurate understanding of the physics within the Sun
(e.g.,\citep{2013MNRAS.435.2109L}). Based on this, we assume that the total mass of the halo must be sufficiently light to have a negligible effect on solar gravity and structure. Furthermore, \citet{2020CmPhy...3....1B} found that planetary ephemerides data can exclude dark matter halos with a total mass greater than $10^{-12}M_\odot$. As a result, we have adopted a total mass of $10^{-13}M_\odot$ for our dark matter halo unless otherwise specified. Our fiducial model assumes $m_\psi=1.5\times 10^{-14}\;{\rm eV}$ and $M_\psi=10^{-13}M_\odot$, which yields $R_\psi=168\;R_\odot=0.783\;{\rm A.U.}$ and $\chi_\psi=5.69 \times 10^{4}$. Occasionally, we consider a larger mass for the dark matter halo, such as $M_\psi=10^{-10}M_\odot$.

\section{Active neutrinos propagating  in a boson dark matter field}
\label{sec-ALDM}

Here, we choose to explore the physics beyond the standard model by encoding  nonstandard interactions between active neutrinos and  dark matter 
\citep{2015PhLB..744...55M}
in the framework of effective field theory (e.g.,\citep{2020RPPh...83l4201A}). We start by considering an extended version of the standard three-neutrino flavour oscillations model \citep{2008PhR...460....1G}, which includes an additional nonstandard interaction: we postulated that the three active neutrinos $\nu_a$ ($a=e,\tau,\mu$) could change their flavour through an interaction with an ultralight time dependent boson field $\psi(r,t)$ [Eq. \eqref{eq:psi2}], mediated by vector boson $\phi$ with a mass  $m_\phi$. 
Therefore, $\psi$ couples with $\nu_a$ by the interaction $g_{\psi} \psi \nu_a\nu_a$, where $g_\psi$ is a dimensionless coupling (e.g.,\citep{2019JHEP...12..046S}).  Moreover, to widen the space of possible solutions, the intermediate boson $\phi$ can be an ultralight particle.
Accordingly,  the effective Lagrangian (e.g.,\citep{2018PhRvD..97g5017K}) that describes the system is
\begin{eqnarray}
	{\cal L} \supset -m_{\nu}\left(1 + g_\psi\frac{\psi}{\Lambda}\right)\nu\nu +{\rm H.c.},
	\label{eq:Lagrangian} 
\end{eqnarray}
where $\Lambda=m_\nu/g_\psi $ is a large mass scale (for instance, with a $ g_\psi \ll 1$). For convenience, we suppress the flavour indices on $\nu$ and $m_\nu$. The results found in this work are equally valid for  Dirac and Majorana neutrinos.
Therefore, the survival probabilities of solar neutrinos can vary through two new mechanisms:
\begin{enumerate}
	\item
	Neutrino masses inside the dark matter halo can change according to [Eq. \eqref{eq:Lagrangian}]: $m_\nu$  can vary to  $m_\nu\left(1+\delta  m_\nu/m_\nu \right)$   by the action of the boson field.  Here we consider a large number of bosons within the de Broglie wavelength, making them oscillate coherently as a single classical field, such as $\psi (r,t)$  corresponds to the boson field defined in [Eq. \eqref{eq:psi2}], for which the mass perturbation reads
	\begin{eqnarray}
		\frac{\delta m_\nu}{m_\nu}=\bar{\epsilon}_\psi e^{-\frac{r}{R_{\psi}}}\cos{\left(m_\psi t\right)},
		\label{eq:mnu}
	\end{eqnarray}
	where the amplitude  $\bar{\epsilon}_\psi$ reads
	\begin{eqnarray}
		\bar{\epsilon}_\psi=
		\frac{\sqrt{m_\psi}}{2}\;
		\frac{g_\psi\;\sqrt{2 \chi_\psi \bar{\rho}^{\odot}_{\rm DM} }}{m_\nu m_\psi}
		\left(\frac{h^2\Omega_{\psi}}{h^2\Omega_{\rm DM}}\right)^{1/2}.
		\label{eq:epsilon}
	\end{eqnarray}
	If not stated otherwise, we assume that $\Omega_{\psi}h^2=\Omega_{\rm DM}h^2$. Note that $\bar{\epsilon}_\psi $ can affect neutrino flavors' transformation even if $\Omega_\psi h^2$ is a small fraction of the dark matter halo. \citet[][and references therein]{2020ApJ...905...22L} discusses the properties of such a dark matter model.
	
	\item
	The forward scattering of active neutrinos $\nu_a$ can change due to the boson field $\psi $ and the intermediator vector boson $\phi $\footnote{	
		We note that the MSW potential resulting from the interaction of neutrinos with a boson field through a fermionic mediator is identical to the interaction of neutrinos with a fermion field through a bosonic mediator \citep{2021JHEP...09..177S}. The two particles switch roles in the MSW potentials. The difference between the two MSW potentials may only appear in higher-order terms.}.
	   Such effect is taken into account by the inclusion of a new term $V_{\psi\nu_a}$ in the matter potential diagonal matrix ${\cal V}$ (e.g.,\citep{1989RvMP...61..937K}).   The function $V_{\psi\nu_a}$  is an effective Wolfenstein potential of $\psi$ particles associated with flavour change due to the propagation of neutrinos inside the dark matter limit medium. Such neutrino flavour oscillation results from the neutrinos' interaction with bosons $\psi$ through $\phi$. This process corresponds to the well-known  Mikheyev-Smirnov-Wolfenstein effect \citep[MSW;][]{1978PhRvD..17.2369W,1985YaFiz..42.1441M}.   
	The $V_{\psi\nu_a}$  inside the boson halo for a generic mediator $\phi$ \citep{2019JHEP...12..046S}, reads
	\begin{eqnarray}
	V_{\psi\nu_a}(r)= g_\psi g_{\nu_a} 
		\left\{	
		\begin{array}{ll}
			\bar{V}_{\psi}^{\rm I} (r)+\bar{V}_{\psi}^{\rm II} (r) &\qquad r  \leq {R_\psi} \\
			\\
			\bar{V}_{\psi}^{\rm III}(r) & \qquad r \geq {R_\psi}
		\end{array}
		\right.
		\label{eq:Viphi}	
	\end{eqnarray} 
where $g_\psi $ and $g_{\nu_a} $ represent the coupling constants of the corresponding particle — boson and active neutrino, associated with  an intermediator particle with mass $m_\phi$ (e.g.,\citep{2015PhLB..744...55M}). 
The radial functions $\bar{V}_{\psi}^{\rm I}(r)$,  $\bar{V}_{\psi}^{\rm II}(r)$ and $\bar{V}_{\psi}^{\rm III}(r)$ are   given by the following expressions: 
	\begin{eqnarray} 
		\bar{V}_{\psi}^{\rm I}(r)=
		\frac{e^{-m_\phi r} }{m_\phi r}
		\int_0^r r^\prime \bar{n}_{\psi}(r^\prime)\sinh{(m_\phi r^\prime) } dr^\prime, 
	\end{eqnarray} 	
	\begin{eqnarray} 
		\bar{V}_{\psi}^{\rm II}(r)= \frac{\sinh{(m_\phi r)}}{ m_\phi r}
		\int_r^{R_\psi} r^\prime \bar{n}_{\psi}(r^\prime)e^{-m_\phi r^\prime } dr^\prime
	\end{eqnarray} 	
	and 
	\begin{eqnarray}
		\bar{V}_{\psi}^{\rm III}(r) =
		\frac{e^{-m_\phi r}}{ m_\phi r}
		\int_0^{R_\psi} r^\prime \bar{n}_{\psi}(r^\prime)\sinh{(m_\phi r^\prime) } dr^\prime.
	\end{eqnarray}   
	where $n_\psi(r)$ is the local density of bosons given by
	\begin{eqnarray}
		n_\psi (r) =\frac{\rho_{\psi}(r)}{m_\psi}
		=\frac{\rho_{c\psi}}{m_\psi}\;
		e^{-\frac{2r}{R_{\psi}}},
		\label{eq:npsi}	
	\end{eqnarray} 
where $\rho_{\psi}(r)$ is given by Eq. \eqref{eq:rhopsi}. 	
\begin{table}
\centerline{
\begin{tabular}{rrrrrrrr}	
\hline\hline
Mod. & { $m_\psi$} & 	{  $m_\phi$}   &
{ $g_{\psi\nu}$} & { $\chi^2_{\nu}$}&  { $\chi^2_{\nu}$ per} & Color   \\
---  & ${\rm eV}$&  ${\rm eV}$ & $G_{F}$ &   --- &   {$d.o.f.$}   &   Curve \\
\hline
$\rm S_{\nu}$  & 	---    & 	---  &	---  &  $2.73$ &  $0.55$ &  red \\ 
$\rm R_{a\nu}$ & $1.5\;10^{-14} $   & 		$10^{-12} $    &	$6\;10^{23} $ &   $2.76$  &  $0.69$  & blue\\ 
$\rm R_{b\nu}$ & $1.5\;10^{-14} $   & 		$10^{-12} $    &	$-5\;10^{24} $ &   $2.58$  &  $0.65$ & green\\ 
\\ 	
$\rm A_{1}$ &$1.5\;10^{-14} $   & 	$10^{-12} $  &	$\mathbf{3\;10^{24}} $&   $2.94$ &  $0.74$  & coral \\
$\rm A_{2}$ &$1.5\;10^{-14} $  & 		$10^{-12} $    &	$\mathbf{1\,10^{25}} $ &    $3.77$ &  $0.94$  & gold \\
$\rm A_{3}$ &$1.5\;10^{-14} $   & 	$10^{-12} $    &	$\mathbf{5\;10^{25}} $ &   $15.44$&  $3.86$ & violet \\	
$\rm A_{4}$ &$1.5\;10^{-14} $   & 	$10^{-12} $   &	$\mathbf{1\;10^{26}} $ &   $27.36$ &  $6.84$ & lime \\	
$\rm A_{5}$ &$1.5\;10^{-14} $   & 	$10^{-12} $    &	$\mathbf{-1\;10^{24}} $ &   $2.68$ &   $0.67$ & aqua \\	
$\rm A_{6}$ &$1.5\;10^{-14} $   & 	$10^{-12} $    &
$\mathbf{-7\;10^{25}} $ &   $13.23$&  $3.31$ &  brown \\
\\
$\rm B_{1}$ &$\mathbf{1.0\;10^{-14}} $    & 	$10^{-12} $   &	$6\;10^{23} $ &   $2.74$ &  $0.69$  & ---   \\
$\rm B_{2}$ &$\mathbf{1.8\;10^{-14}} $    & 	$10^{-12} $   &	$6\;10^{23} $ &   $2.80$ &  $0.70$  & ---   \\
\\
$\rm C_{1}$ & $1.5\;10^{-14} $    & 		$\mathbf{10^{-20}} $   &	$6\;10^{23} $  &   $2.73$ &  $0.68$  & ---  \\
$\rm C_{2}$ &  $1.5\;10^{-14} $ & 		$\mathbf{10^{-10}}$     &	$1\;10^{25} $ &   $2.76$ &  $0.69$  & ---  \\
$\rm C_{3}$ &  $1.5\;10^{-14} $   & 		$\mathbf{10^{-5}}$     &	$1\;10^{25} $ &   $2.76$ &  $0.69$  & --- \\
$\rm C_{4}$ &  $1.5\;10^{-14} $   & 		$\mathbf{10^{5}}$  &	$1\;10^{25} $ &   $2.73$ &  $0.68$  & ---  \\
		\\
$\rm D_{1}$ & $1.5\;10^{-14} $   & 	$10^{-12} $    &	$6\;10^{23} $ &   $2.62$ &  $0.66$  & --- \\
$\rm D_{2}$ & $1.5\;10^{-14} $   & 		$10^{-12} $    &	$6\;10^{23} $ &   $2.68 $ &  $0.67$  & --- \\
$\rm D_{3}$ & $1.5\;10^{-14} $   & 		$10^{-12} $      &	$6\;10^{23} $   &   $3.47$&  $0.87$   &--- \\
$\rm D_{4}$ & $1.5\;10^{-14} $   & 		$10^{-12}  $    &	$6\;10^{23} $  &   $4.94$  &  $1.23$ & --- \\
\hline	
\end{tabular} $\qquad\qquad$ }
\caption{Comparison of the parameters of various dark matter boson-neutrino models. The standard three-neutrino model $S_\nu$ is compared to parameter-varying models, including $R_{a\nu}$, $R_{b\nu}$, $A_i$, $B_i$, and $C_i$, where $\bar{\epsilon}=10^{-3}$ is fixed, and $D_i$, where $\bar{\epsilon}=j \times 10^{-1}$ (with $j=1,2,3,4$) is chosen. To highlight the varying parameters across each set of models, the corresponding values of these parameters are denoted in bold. Figures \ref{fig:neff},  \ref{fig:PeB8a} and \ref{fig:PeB8b}  show the effective resonance density $n_{\rm eff}$ and the electron neutrino survival probability $\langle P_{ee}(E)\rangle$ for some models. These figures provide a comparison of the various models discussed in the article. The Sun is assumed to be inside a boson cloud with a mass of approximately $10^{-13}\;M_\odot$, a radius of $R_\psi=0.783\;{\rm A.U.}$, and a parameter of condensation $\chi_\psi=5.7\times10^{4}$.}
\label{tab:neutchi2}
\vspace{0.5cm}
\end{table}

	\begin{figure}[!t]  
		\centering 
		\includegraphics[scale=0.45]{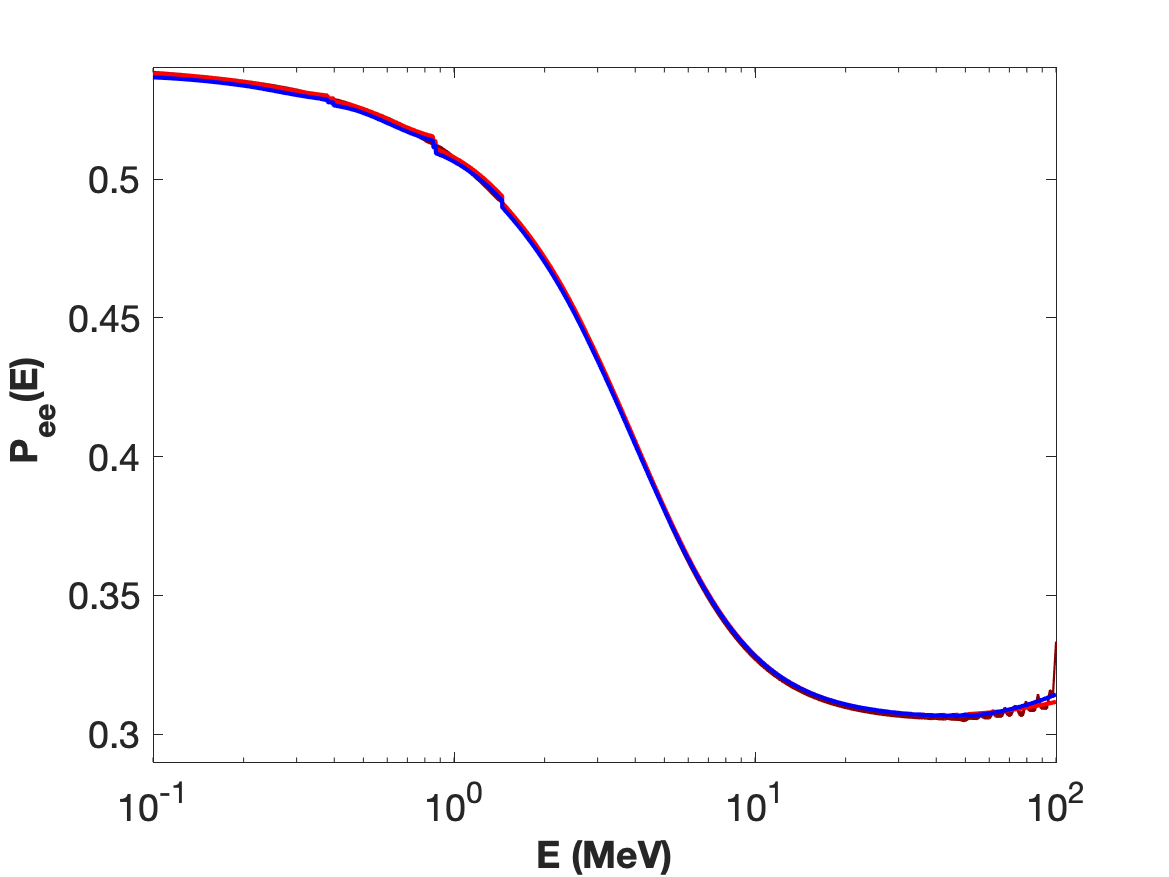}
		\caption{ This figure illustrates the survival probability of the electron neutrino $P_{ee}(E)$ computed for the standard model of neutrino flavor oscillations (model $\rm S_{\nu}$, as detailed in Table \ref{tab:neutchi2}). The calculations, based on Eqs. \eqref{eq:Pee} and \eqref{eq:Pee2}, also factor in the jump probability term, $P_\gamma$ Eq. \eqref{eq:Pgamma}. The red curve includes the $P_\gamma$ contribution, as per  Eq. \eqref{eq:Pgamma}), while the blue curve depicts $P_{ee}(E)$ without the $P_\gamma$ contribution. While the contribution of $P_\gamma$ to $P_{ee}(E)$ remains negligible within the shown neutrino energy interval, its influence becomes marginally visible for neutrino energies exceeding 50 MeV.	 For more details, please refer to the main text.}
		\label{fig:Peeeff}
	\end{figure}
	
\begin{figure}[!t]  
	\centering 
	\includegraphics[scale=0.45]{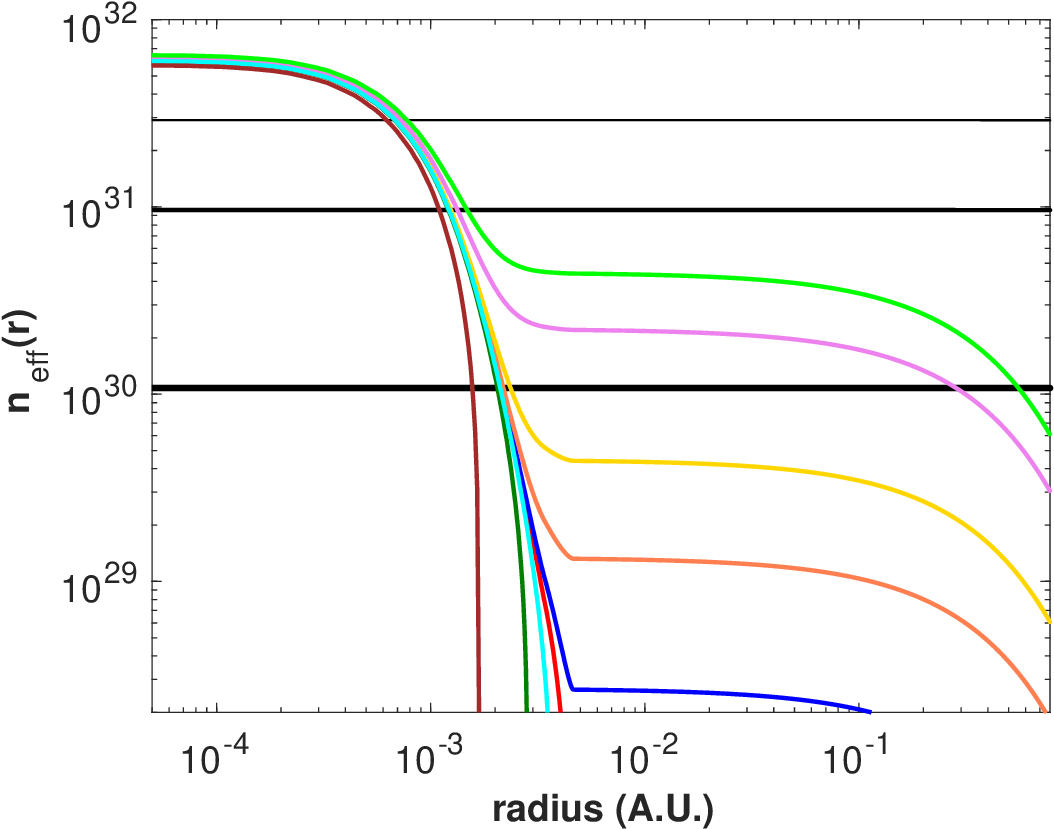}
	\caption{The effective density $n_{\rm eff}(r)$ is shown as a function of radius within a spherical boson halo with a radius of $R_\psi=0.783\;{\rm A.U.}$ and a total mass of $M_\psi\approx 10^{-13}\;M_\odot$, composed of bosons with a mass of $m_\psi=1.5\times 10^{-14}\; {\rm eV}$. The {\bf red curve} represents the standard three-neutrino flavor oscillation model $\rm S_{\nu}$ (see Table \ref{tab:neutchi2}), where $n_e(r)=n_{\rm res}(E)$ [Eq. \eqref{eq:Nres}]. The horizontal black lines correspond to $n_{\rm res}(E)$ for $E$ values of 100, 11, and 4 MeV, which occur in the layer located at 0.46, 0.26, and 0.15 of the solar radius, respectively (see main text for details). In our model, active neutrinos interact with the bosons through an intermediary particle with mass $m_\phi$ and a coupling constant $g_{\psi\nu}$. The other curves shown use the condition $n_{\rm eff}(r)=n_{\rm res}(E)$ [Eq. \eqref{eq:Nresnew}] for $E=100$ MeV. These curves correspond to the models $\rm R_{a\nu}$, $\rm R_{b\nu}$ and $\rm A_i$ ($i=1,6$) presented in Table \ref{tab:neutchi2}, with the color scheme indicated in the table.}
	\label{fig:neff}
\end{figure}
	
It follows  that the Wolfenstein potential 
$V_{\psi\nu_a}$ [Eq. \eqref{eq:Viphi}] now reads
	\begin{eqnarray}
		V_{\psi\nu_a} (r)=\frac{g_\psi g_{\nu_a} R_\psi^2 }{m_\phi^2R_\psi^2-4}\;n_{\psi} (r)\; [1+ \Psi(r)].
		\label{eq:Vepn-e}
	\end{eqnarray}
	The coupling $g_{\nu_a} $ is different for neutrino with different flavours: $\nu_a$ ($a=e,\nu,\tau$).
	The function
	$ \Psi(r)$ is given by
	\begin{eqnarray}
		\Psi(r)=\frac{4}{m_\phi^2R_\psi^2-4}\frac{R_\psi}{r}\left[ e^{-(m_\phi R_\psi-2)\frac{r}{R_\psi}}-1\right].\;\;\,
		\label{eq:Phiepn-e}
	\end{eqnarray}
	This result corresponds to the effective potential  of spherically symmetric
	exponential density distribution  \citep{2019JHEP...12..046S}. If $R_\psi $ 
	is much larger, then
	the effective potential corresponds to a pointlike interaction: $V_{\psi\nu_a} =g_\psi g_{\nu_a} n_{\psi}/m_\phi^2$. We remind the reader that $\nu$ can be one of the following flavours:
	$\nu_e$, $\nu_\tau$ and $\nu_\mu$.
	
\end{enumerate}

\medskip\noindent
We remind the reader that the interaction between dark matter particles with neutrinos and antineutrinos depends intrinsically on the nature of the dark matter particle and the particle mediator.  
 n this work, without loss of generality, we assume that the dark matter interacts with neutrinos but not with antineutrinos. 
It is worth reminding the reader that such asymmetry is already present in the standard MSW effect (e.g.,\citep{1978PhRvD..17.2369W}) since most of the neutrino propagation medium is composed of matter and not antimatter.   Moreover, we also include in our calculation the corrections resulting from the propagation (at finite temperature) of neutrinos in a thermal background of dark matter particles (e.g.,\citep{1987PhRvD..35..896B,1992PhRvD..46.1172D}).
Several mechanisms contribute to the enhancement or suppression of the conversation of neutrinos from one flavour to another; however, some of these processes are more relevant than others for the neutrino energy window of this study. 
Here, following   \citet{2000NuPhB.583..260L},   we include this effect using the effective propagator\footnote{
We note that for convenience, the definition of the propagator function in this work differs from the one found in the literature \citep{2000NuPhB.583..260L,2006PhLB..634..267K}, where 
	$\zeta_o(s_\phi)={(1-s_\phi)}/{[(1-s_\phi)^2+\gamma_\phi^2]}$, therefore $\zeta_\phi=\zeta_o(-s_\phi)$.} of the intermediator $\phi$
defined as 
$
\zeta_\phi={(1+s_\phi)}/{[(1+s_\phi)^2+\gamma_\phi^2]}$  where  $s_\phi=2 Em_\psi/m_\phi^2$,
$\gamma_\phi=\Gamma_\phi/m_\phi$ and $\Gamma_\phi$ is the width of the intermediator particle $\phi$.  
\citet{2006PhLB..634..267K} 
have computed the propagator function using the thermal field theory
\citep{1988NuPhB.307..924N}  and found that the propagator  function should read
 $ \zeta_\phi={(1+s_\phi)}/{[(1+s_\phi)^2+s_\phi^2\gamma_\phi^2]} $.  This last $\zeta_\phi$ expression differs from the expression found by \citet{2000NuPhB.583..260L}, only by a term of small magnitude in the dominator of function $\zeta_\phi$,  where $s_\phi^2\gamma_\phi^2$  replaces $\gamma_\phi^2$.
 Since $\gamma_\phi$  is a minimal quantity, both expressions agree with current experiments' data \citep{2000NuPhB.583..260L}. 
Therefore, here, we opt for the following expression: 
\begin{eqnarray}
\zeta_\phi\equiv \frac{(1+s_\phi)}{(1+s_\phi)^2+\gamma_\phi^2}.	\label{eq:zetaphi}
\end{eqnarray}
It is worth noticing that in the limit of $s_\phi\gg 1$, we obtain  
$\zeta_\phi \approx s_\phi^{-1}=m_\phi^2/(2Em_\psi) $. 
We include the  correction on  $V_{\psi\nu_a} $ [Eq. \eqref{eq:Vepn-e}]), thought the function $\zeta_\phi$ [Eq. \eqref{eq:zetaphi}], hence the Wolfenstein potential, now reads
\begin{eqnarray}
	V_{\psi\nu_a} (r)=\frac{g_\psi g_{\nu_a} R_\psi^2 }{m_\phi^2R_\psi^2-4}\;n_{\psi} (r)\; \zeta_\phi\; [1+ \Psi(r)].
	\label{eq:Vepn-e2}
\end{eqnarray}
We interpret the function $\zeta_\phi$  in $V_{\psi\nu_a} (r)$ as the correction coming from the propagation of neutrinos within an effective bosonic-neutrino potential that starts by considering the effect of neutrino propagation on a limited bosonic medium.

\section{Survival probability of electron neutrinos: classical model}
\label{sec-PLDM}
Here, we compute the survival probability of electron neutrinos $P_{ee}(E)$ of several ultralightdark matter models and compare them with the data coming from solar neutrino detectors.
We compute $P_{ee}(E)$  by using one of the analytical formulas found in the literature (e.g.,\citep{1986PhRvL..57.1271H,1986PhRvL..57.1275P,2013PhRvD..88d5006L,2013ARA&A..51...21H,2017ChPhC..41b3002B}). 
If not stated otherwise, we consider the neutrino oscillations occurring inside the Sun and in the boson halo to be adiabatic. A detailed discussion about adiabatic and nonadiabatic neutrino flavour oscillations is available in  \citet{2003RvMP...75..345G,2018arXiv180205781F}.
In agreement with the current neutrino data in which, at a good approximation, the neutrino flavour oscillations between the three flavours are considered adiabatic \citep{2004NJPh....6...63B,2017ChPhC..41b3002B,2021Univ....7..231K}, and following the recent review of particle physics on this topic \citep{2018PhRvD..98c0001T,2016ChPhC..40j0001P}, 
specifically the article  "Neutrino Masses, Mixing, and Oscillations", the survival probability of electron neutrinos reads
\begin{eqnarray}
	P_{ee}(E)\approx\cos^4{(\theta_{13})}P_{ee}^{2\nu_e}+\sin^4{(\theta_{13})}
	\label{eq:Pee}	
\end{eqnarray}
where $P_{2\nu_e}$ is the survival probability in the two neutrino flavour model (with $\theta_{13}=0$) given by 
\begin{eqnarray}
	P_{ee}^{2\nu_e}(E)=\frac{1}{2}+\left(\frac{1}{2}-P_{\gamma}\right)\cos{(2\theta_{12})} 
	\cos{(2\theta_{m})} 
	\label{eq:Pee2}		
\end{eqnarray}
where $P_\gamma$ is the jump probability that corrects the adiabatic expression (\ref{eq:Pee2}) for the nonadiabatic contribution, 
and $\theta_{m}=\theta_{m}(r_s)$ is the matter mixing angle at the point of neutrino production (source)  located at a distance  $r_s$ from the center of the Sun. (e.g.,\citep{1989PhRvD..39.1930K,1995PhRvD..51.4028B}). The  jump probability $P_\gamma$ \citep{2003NIMPA.503....4D}   reads  
\begin{eqnarray}
	P_\gamma=\frac{e^{-\gamma\sin^2{\theta_{12}}}-e^{-\gamma}}{1-e^{-\gamma}} P_{\rm H}
	\label{eq:Pgamma}	
\end{eqnarray}
where $\gamma=2\pi h_\gamma \Delta m_{21}^2/2E$, $h_\gamma$ is the scale height \citep{2000PhLB..490..125G} and $P_{\rm H}$ is a regular step function.

The  matter mixing angle  (\citet{2011PhRvD..83e2002G}) is  given by
\begin{eqnarray}
	\cos(2\theta_{m})=\frac{A_m}{
		\sqrt{A_m^2 +\sin^2{(2\theta_{12})}  }}
	\label{eq:sintheta12}	
\end{eqnarray}
where $A_m$  reads
\begin{eqnarray}
	A_m=\cos{(2\theta_{12})}-{V_m}/{\Delta m^2_{21}}. 
	\label{eq:Am}	
\end{eqnarray}
In the standard case \citep{2004NJPh....6...63B},  it corresponds to
$V_{m}=2V_{cc}\cos^2{(\theta_{13})}E$ with $V_{cc}= \sqrt{2}G_{F}n_e(r) $ where $n_e(r)$ is different from zero ($r\le R_\odot$) only inside the Sun. In the previous expression, $V_{cc}$ is the Wolfenstein potential.  Nevertheless, as we will see in the next section, $V_{cc}(r)$  in this study will be replaced by a new effective potential $V_{\rm eff}(r)$.

\medskip\noindent 
The maximum production of neutrinos in the Sun's core occurs in a region between 0.01 and 0.25 solar radius, with neutrino nuclear reactions of the proton-proton chain and carbon-nitrogen-oxygen cycle occurring at different locations (e.g.,\citep{2013ApJ...765...14L}). These neutrinos, produced at various values of $r_s$, when traveling towards the Sun's surface, follow paths of different lengths. Moreover, during their travelling, neutrinos experience varying plasma conditions, including rapid decreasing of the electron density from the center towards the surface. In general, we expect that nonadiabatic corrections averaged out and are negligible along the trajectory of the neutrinos, except at the boundaries (layer of rapid potential transition) of the neutrino path, typically around the neutrino production point or at the surface of the Sun\footnote{Since the potential is zero at the Sun's surface, the nonadiabatic contribution is negligible.}.
Therefore, we could expect Eq. \eqref{eq:Pee2}  to be very different when taking such effects into account. Nevertheless, this is not the case, 
\citet{2004NuPhB.702..307D} analyzed in detail the contribution to $P_{ee}$
[Eq. \eqref{eq:Pee}]  coming from nonadiabaticity corrections and variation on the locations of neutrino production, i.e., $r_s$, and they found that the impact is minimal. In general, $P_\gamma$ [Eq. \eqref{eq:Pgamma}] is expected to take a real value, such that  $P_\gamma=0$  or $P_\gamma \ne 0$  corresponding to
neutrino flavour adiabatic and nonadiabatic conversions. In general, the conversions are only called nonadiabatic if $P_\gamma$ is non-negligible. Figure \ref{fig:Peeeff} depicts the survival probability $P_{ee} (E)$ for the standard three-neutrino flavour oscillation model. Interestingly, the contribution of $P_\gamma$ to $P_{ee}(E)$ is minimal, bordering on negligible. In this figure, the red and blue curves represent $P_{ee}(E)$, as defined by equation (\ref{eq:Pee2}), with and without the $P_\gamma$ contribution, respectively, the latter being determined by Eq. \eqref{eq:Pgamma}.

\medskip\noindent
Since the electron number density $n_e$ varies considerably along
the neutrino path in the Sun: $n_e$ decreases monotonically from the
$10^{31}\; m^{-3}$  in the center of the star to an almost negligible value at the surface, e.g.,\citep{2013ApJ...765...14L}. Therefore,  neutrinos propagating toward the surface necessarily cross a layer of matter where $n_e=n_{\rm res}$ such that $A_m=0$. This particular solution of the function $A_m$, the value for which $n_{\rm res}(E)$ leads to $A_m=0$ is known as the
resonance condition. In the classic case, we compute this electronic density associated with the  resonance condition as
\begin{eqnarray}
	n_e(r)=	n_{\rm res} (E)\equiv	
	\frac{\Delta m^2_{21} }{2\sqrt{2}G_{F}E}
	\frac{\cos{(2\theta_{12})}}{\cos^2{(\theta_{13})}}.
	\label{eq:Nres}
\end{eqnarray}
where $r= r_\gamma$ ($\ne h_\gamma$) is defined as the layer where  the resonance condition $n_e(r)=	n_{\rm res} (E)$ occurs. 
Figure \ref{fig:neff} shows $n_{\rm res}(r) $ for the present Sun as a continuous red curve. In the same figure, the horizontal lines correspond to energy values of electron neutrino equal to 100, 11 and 4 MeV, for which the resonance condition occurs for the radius of 0.46, 0.26 and 0.15 solar radius.  

\medskip\noindent
In general,  the adiabatic and nonadiabatic nature of neutrino oscillations depends on the neutrino's energy $E$ and the relative value of the resonance condition
of  $n_{\rm res}(E)$ [Eq. \eqref{eq:Nres}]. For instance,  if a neutrino of energy $E$ is such that (i) $n_{\rm res} (E) \gg n_{\rm e}$ neutrinos oscillate practically as in vacuum; (ii) $n_{\rm res } (E)\ll n_{\rm e}$ oscillations are suppressed in the presence of matter \citep{2016ChPhC..40j0001P}.

\medskip\noindent
In our models, most cases correspond to adiabatic transitions, for which $P_\gamma\approx 0$.  Nevertheless, it is possible to compute the contribution of the nonadiabatic component $P_\gamma$ to $P_{ee}(E)$ by using Eq. \eqref{eq:Pgamma} and the following prescription: (i) compute the value of $n_{\rm res}$  [using Eq. \ref{eq:Nres}] for each value of $E$ (with fixed values of $\Delta m_{12}^2$, $\theta_{12}$ and $\theta_{13}$), (ii) calculate the scale height $h_\gamma =|n_e/(d n_e/dr)|_{r_\gamma}$ at the point $r_\gamma$ defined as $n_e(r_\gamma)=n_{\rm res}(E)$, and (iii) calculate $P_\gamma$ and $\gamma$ for the value of $h_\gamma$. The scale height $h_\gamma$ also reads $h_\gamma =|(d \ln{n_e}/dr)^{-1}|_{r_\gamma}$,  the reason for which  will be made clear later. 
We also found that  $h_\gamma=|(d \ln{V_{cc}}/dr)^{-1}|_{r_\gamma}$.

\medskip\noindent 
Conveniently, to properly take into account the nonadiabatic correction into Eqs. \eqref{eq:Pee2} and \eqref{eq:Pgamma}, we included the step function  $P_{\rm H}$, defined as $P_{\rm H} (V_m - \Delta m^2_{21}   \cos{(2\theta_{12})} )$. This function is one for
$ \Delta m^2_{21}   \cos{(2\theta_{12})}\le V_m $, and is 0 otherwise
(e.g.,\citep{2000PhRvD..61j5004C}).

\begin{figure}[!t] 
\centering 
\includegraphics[scale=0.47]{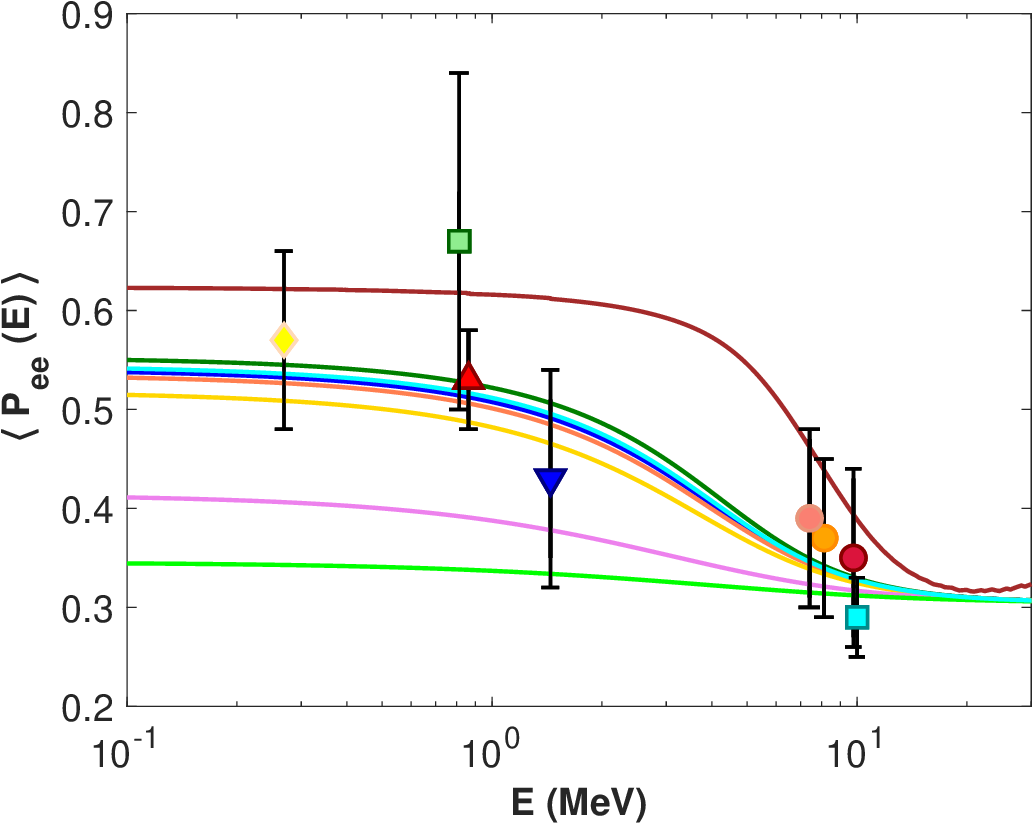}	\caption{
Survival probability of electron neutrinos $\langle P_{ee}(E)\rangle$ in a standard three-neutrino flavour oscillation model with neutrinos, coupled to a boson dark matter field $\psi$ with a fixed value of $m_\psi$ (see main text). The figure shows the $\langle P_{ee}(E)\rangle$ of boson models presented in Table \ref{tab:neutchi2} with $\bar{\epsilon}=10^{-3}$. The curves correspond to a set of models presented in Table \ref{tab:neutchi2}: $\rm S_{\nu}$ (red curve, ${\chi_\nu}^2=2.73$), $\rm R_{a\nu}$ (blue curve, ${\chi_\nu}^2=2.76$), $\rm R_{b\nu}$ (green curve, ${\chi_\nu}^2=2.58$), and $\rm A_i$ models, for instance, $\rm A_2$ (gold curve, ${\chi_\nu}^2=3.77$), $\rm A_3$ (violet curve, ${\chi_\nu}^2=15.44$), and $\rm A_6$ (light brown curve, ${\chi_\nu}^2=13.23$). The data points correspond to the survival probabilities of electron neutrinos measured by three solar neutrino detectors (SNO, Super-Kamiokande, and Borexino) and computed using a current standard solar model: (i) Borexino: $pp$ (yellow diamond), $^7Be$ (red upward triangle), $pep$ (blue downward triangle), and $^8B$ HER (salmon circle), $^8B$ HER-I (orange circle), and $^8B$ HER-II (magenta circle); (ii) SNO: $^8B$ (cyan square); (iii) KamLAND/SNO: $^7Be$ (green square). See \citet{2018Natur.562..505B,2019PhRvD.100h2004A,2010PhRvD..82c3006B,2011PhRvC..84c5804A,2016PhRvD..94e2010A,2013PhRvC..88b5501A,2008PhRvD..78c2002C}, and references therein, for details on the experimental data. The acronym HER stands for High-Energy Region. See the main text for more details.}
	\label{fig:PeB8a}
\end{figure}

\section{Survival probability of electron neutrinos: new model}
\label{sec-PLDM2}

\medskip\noindent
The survival probability of electron neutrinos [Eq. \eqref{eq:Pee}] in this study can vary in comparison to the standard three-neutrino flavour oscillation model by two mechanisms:  
\begin{enumerate}
	\item
	Variation of the mass-square differences $\Delta m^2_{ij}$ and the mixing angles $\theta_{ij}$ (where $i,j=1,2,3$ and $i\ne j$), related with the neutrino mass correction $\delta m_\nu$ resulting from the interaction of active neutrinos with the boson field $\psi$:
	\begin{itemize}
		\item[—] 
		Considering only first-order perturbation, thus, the neutrino mass-squared difference reads
		\begin{eqnarray}
			\frac{\Delta m_{ij}^2(r,t)}{ \Delta m^2_{ij,o}}\approx
			1+2\bar{\epsilon}_\psi e^{-\frac{r}{R_\psi}} \cos{(m_\psi t)} 
			\label{eq:deltamphi}
		\end{eqnarray}
		where $\Delta m^2_{ij,o}=m_i^2-m_j^2$ is the standard (undistorted) value and $\Delta m_{ij}^2(r,t)$ the perturbed term.
		$\Delta m_{ij}^2(r,t)$ varies with  the amplitude $\bar{\epsilon}_\psi (g_\psi,m_\psi)$ and the frequency $m_\psi$.
		In the derivation of Eq. \eqref{eq:deltamphi}, we consider that $\delta m_i/m_i\approx \delta m_\nu/m_\nu$.
		$ \delta m_\nu/m_\nu$ is given by Eq. \eqref{eq:mnu} and 
		$\bar{\epsilon}_\psi $ by  Eq. \eqref{eq:epsilon}.
		The mass-squared difference $\Delta m^2_{ij}$  between neutrinos of different flavours follows the usual convection (e.g.,\citep{2017PhRvD..95a5023L}) such  that $\Delta m^2_{i1}=m^2_{i}-m^2_{1}$ ($i=2$, and $3$). 
		
		\item[—] 
		Similarly, the mixing angle variations is written as
		\begin{eqnarray}
			\theta_{ij}(r,t)- 
			\theta_{ij,o}\approx \bar{\epsilon}_\psi e^{-\frac{r}{R_\psi}}   \cos{(m_\psi t)},
			\label{eq:thetaphi}
		\end{eqnarray} 
		where $\theta_{ij,o}$ is the standard (undistorted) mixing angle. The indices 
		$i$ and $j$ in $\theta_{ij}$ follow a convention identical to the mass-squared differences. 
	\end{itemize}
	Equations  (\ref{eq:deltamphi}) and (\ref{eq:thetaphi}) are time-dependent variations similar to the ones found by several authors, such as \citet{2018PhRvD..97g5017K} and \citet{2016PhRvL.117w1801B}, that result from the impact of $\psi$ on neutrino flavour oscillations. However, in our case, both quantities also vary with the distance.  
	
	\item
	The forward scattering of active neutrinos on the boson dark matter field $\psi$  is taken into account by the inclusion of the new term $V_{\psi\nu_a} $ [Eq. \eqref{eq:Vepn-e2}] in the matter potential diagonal matrix
	${\cal V}={\rm diag}(V_{cc}+V_{nc}+V_{\psi\nu_e},V_{nc}+V_{\psi\nu_\mu},V_{nc}+V_{\psi\nu_\tau})$ where $V_{cc}=\sqrt{2} G_F\; n_e(r)$ and  $ V_{nc}=-G_F/\sqrt{2}\; G_F n_n(r)$  correspond to the charged current (cc) that takes into account the forward scattering of $\nu_e$ with electrons and the neutral current (nc) related to the scattering of all active neutrinos with the ordinary fermions (e.g.,\citep{2020PhR...854....1X}).  Now, if we consider that 
	$V_{\psi\nu_a}$ is the same for all active neutrinos  then the 
	diagonal matrix \citep{1989RvMP...61..937K}  takes the standard form matter potential
	${\cal V}={\rm diag}(V_{cc},0,0)$.  Nevertheless, in the study we consider  $g_{\nu_e}\ne g_{\nu_\mu}=g_{\nu_\tau}$,  therefore   ${\cal V}={\rm diag}(V_{\rm eff},0,0)$ where  $V_{\rm eff}=V_{cc}+V_{\psi\nu_e}-V_{\psi\nu_\mu}$.
	
\end{enumerate}

\medskip\noindent
In the standard model, matter potential ${\cal V}$ associated with the forward scattering of active neutrinos depends strongly on the properties of constitutive particles of the background medium \citep{1987PhRvD..35..896B,1988NuPhB.307..924N}. Therefore, we opt to consider that these neutrinos propagate in the boson medium \citep{2004NuPhB.683..219B,2007PhRvD..75k5017F,2015PhLB..744...55M,2018PhLB..782..641I,2020PDU....3000606C}, for which the coupling constants have the following relations: 
$g_{\nu_e}\ne g_{\nu_\tau}=g_{\nu_\mu}$. 
We  notice that the coupling  of active neutrinos with dark matter background (including axions) have been studied in many scenarios; among other articles, see the following ones:
\citet{1995PhRvD..52.6607B,1996PhLB..375...26B,2004NuPhB.683..219B,2006PhRvD..74d3517M,2007PhRvD..75k5017F,2012PhRvL.109w1301V}. Therefore, now
$V_{\rm eff}$  reads
\begin{eqnarray}
	V_{\rm eff}(r)=\sqrt{2}G_F\left[ n_e+ 
	\frac{ g_{\psi\nu}R_\psi^2 n_{\psi} \zeta_\phi }{m_\phi^2R_\psi^2-4}\left[1+ \Psi(r)\right]
	\right],
	\label{eq:Veff}		 
\end{eqnarray} 
where $g_{\psi\nu}$ is a coupling constant given by
$g_{\psi\nu}=g_\psi (g_{\nu_e}-g_{\nu_\mu})/(\sqrt{2}G_F)$, and $n_e(r)$ is different from  zero only inside the Sun (for $r\le R_\odot$).  Since  $g_{\nu_e}$ and $g_{\nu_\mu}$
are two free positive parameters of different magnitudes, it follows from the definition that $g_{\psi\nu}$ can be positive or negative.  
Accordingly, $V_m$ in Eqs. \eqref{eq:sintheta12} and \eqref{eq:Am}	now reads,  $V_{m}=V_{\rm eff}\cos^2{(\theta_{13})}E$. 

\medskip\noindent
In the new model, we generalize the result given by
Eq. \eqref{eq:Nres} by introducing the
effective density $n_{\rm eff}$ associated with the resonance condition  as
\begin{eqnarray}
	n_{\rm eff}(r_\gamma)=	n_{\rm res}(E)\equiv\frac{\Delta m^2_{21} }{2\sqrt{2}G_{F}E}
	\frac{\cos{(2\theta_{12})}}{\cos^2{(\theta_{13})}}.
	\label{eq:Nresnew}
\end{eqnarray}
where $r_\gamma$ ($\ne h_\gamma$) is defined as the layer where  the resonance condition occurs, and 
$V_{\rm eff}=\sqrt{2}G_F n_{\rm eff} $ [see Eqs \eqref{eq:Veff}], such that $n_{\rm eff}$ reads
\begin{eqnarray}
	n_{\rm eff}=n_e \left[ 1+\frac{n_\psi}{n_e} \zeta_\phi
	\frac{g_{\psi\nu}R_\psi^2}{m_\phi^2R_\psi^2-4}\left[1+ \Psi(r)\right] \right].		
	\label{eq:neff} 	 		  
\end{eqnarray}
Once we assume that the second term of Eq. \eqref{eq:neff} is always positive,
we therefore  choose a boson such that $m_{\phi}\ge m_{\phi,crit}=2/R_\psi$,  the conclusions found at the end of section \ref{sec-PLDM} associated with the electronic density $n_e(r)$ remain valid for $n_{\rm eff} (r)$. For example for $R_\phi=0.5 \; A.U.$ we obtain $m_{\phi,c}= 5.3 \times 10^{-27}\;{\rm GeV}$.
Figure  \ref{fig:neff} shows $n_{\rm eff}(r) $ [Eq. \eqref{eq:neff}] of several models for the present Sun as coloured curves.
Table \ref{tab:neutchi2} presents the parameters of such models.
Models with changed parameters are displayed in bold to aid the reader's understanding of the table.
 It is worth noticing that in comparison with the classical case (equation \ref{eq:Nres}), neutrino oscillations in some of these new models are suppressed for higher neutrino energy values  (equation \ref{eq:Nresnew}). This is the case of models $A_2$  (gold curve) and $A_3$ (violet curve).

\medskip\noindent
Similar to the standard case (model $S_\nu$ in table \ref{tab:neutchi2}), $P_{ee}(E)$'s contribution [Eqs. \eqref{eq:Pee} and \eqref{eq:Pee2}]   coming from the jump probability $P_\gamma$ although small, is not entirely negligible in this class of models.    We compute the jump probability $P_\gamma$  by using expression  (\ref{eq:Pgamma}) where $r_\gamma$ is now determined by condition (\ref{eq:Nresnew})  and the scale height reads $h_\gamma=|(d \ln{V_{\rm eff}}/dr)^{-1}|_{r_\gamma}$. 
The contribution comes from the interaction of electron neutrinos with the boson field $\phi$ within the dark matter halo. Unlike in the standard case (model $S_\nu$ in table \ref{tab:neutchi2}), the contribution is small but marginally visible for the conversion of electron neutrinos with high energy (see Fig. \ref{fig:Peeeff}).

\section{Light Dark Matter Impact on Solar Neutrinos}
\label{sec-LDMeneutrinos}

The survival probability of electron neutrinos $P_{ee}(E)$ [Eq. \eqref{eq:Pee}] is a time-dependent function through the time-varying boson field $\psi(t)$.  Conveniently, we define the averaged   survival probability of electron neutrinos as 
\begin{eqnarray}
	\langle P_{ee}(E) \rangle=\int_0^{\tau_\psi} P_{ee} (E,\psi)	\frac{dt}{\tau_\psi},
	\label{eq:Peet}
\end{eqnarray}
where   $\tau_\psi=2\pi/m_\psi$ is the period of the boson field $\psi(t)$. 
The ability of a solar neutrino detector to measure the impact of the time-dependent field $\psi(t)$ on the averaged survival probability  $\langle P_{ee}(E) \rangle$ depends on three characteristic timescales: 
\begin{enumerate}
	\item	
	$\tau_\nu$ is the neutrino flight time, for a solar neutrino $\tau_\nu$   is approximately  8.2 min; 
	\item
	$\tau_{ev}$ is the time between two consecutive neutrino detections; for some of the forthcoming neutrinos experiments, $\tau_{ev}$  is bigger than 7 min
	\citep[JUNO,][]{2015arXiv150807166A}.  
	\item
	$\tau_{ex}$ is the total run time of the experiment, which for most detectors should be above ten years. 
\end{enumerate}
Since solar neutrino detectors will run for long periods and collect many events, it is reasonable to consider that  $\tau_{ev}$ and $\tau_{ex}$  have small and large values, respectively. Therefore, the  $\langle P_{ee}(E) \rangle$ time modulation by $\psi(t)$ depends slowly on period $\tau_\psi$ in comparison to  $\tau_\nu$.
Hence, from the condition that $\tau_\psi=8.2\; min=\tau_\nu$, we obtain a critical boson mass $m_{\psi,c}=8.3\;10^{-18}\; eV$. This critical value defines the mass range of the two time modulation regimes that affect the survival probability function of the electron neutrinos:
\begin{enumerate}	
	\item{\bf Low-frequency regime:}	
	$ m_{\psi} \le m_{\psi,c} $ (or $ \tau_{\psi} \ge \tau_{\nu} $), a direct  time modulation of  $ P_{ee} (E,\psi)$ occurs when the period of $\psi(t)$ is  larger than
	the neutrino flight time $\tau_{\nu}$. In this case a temporal variation of the neutrino signal may be observed in the $\langle P_{ee}(E) \rangle$ function. 
	This type of physical process and the associated variation of time-dependent neutrino flux measurements were studied by \citet{2016PhRvL.117w1801B}, among others.

	\item{\bf High-frequency regime:}	
	$ m_{\psi} \ge m_{\psi,c} $ (or $ \tau_{\psi} \le \tau_{\nu} $),
	the change of $ P_{ee} (E,\psi)$  produced by $\psi(t)$ occurs in a time scale faster than the neutrino flight time $\tau_\nu$. 
	The timescale of this variation is too quick to create a  periodic time modulation on the neutrino flux measurements. 
	Nevertheless, such a process leads to the existence of a distorted $\langle P_{ee}(E) \rangle$ average and a spread of the $P_{ee}(E,\psi)$, similar to an energy resolution smearing. This distorted probability average is identified by its deviation from the undistorted  $\langle P_{ee}(E) \rangle$ in the standard scenario \citep[e.g.,][]{2018PhRvD..97g5017K}. Indeed,  the net effect of averaging over time [see Eq. \eqref{eq:Peet}] induces a shift in the observed values of  $\langle P_{ee}(E) \rangle$ relative to its undistorted value.
\end{enumerate}

\medskip\noindent
The neutrino models discussed in this work are within the later case — {\it the high-frequency regime}, once  $ m_{\psi}$ (see table \ref{tab:neutchi2})  is much larger than  $ m_{\psi,c} $ for all the models.  It is worth highlighting that future neutrino detectors will obtain neutrino flux datasets that we can use to test such a range of boson masses. Examples of such class of detectors are the Deep Underground Neutrino Experiment \citep[DUNE,][]{2015arXiv151206148D}   and the Jiangmen Underground Neutrino Observatory  \citep[JUNO,][]{2016JPhG...43c0401A}.

\section{Testing ultralightbosons with active neutrinos}
\label{sec-LDMeneutrinos2}
\begin{figure}[!t]
\centering 
\centering{\includegraphics[scale=0.45]{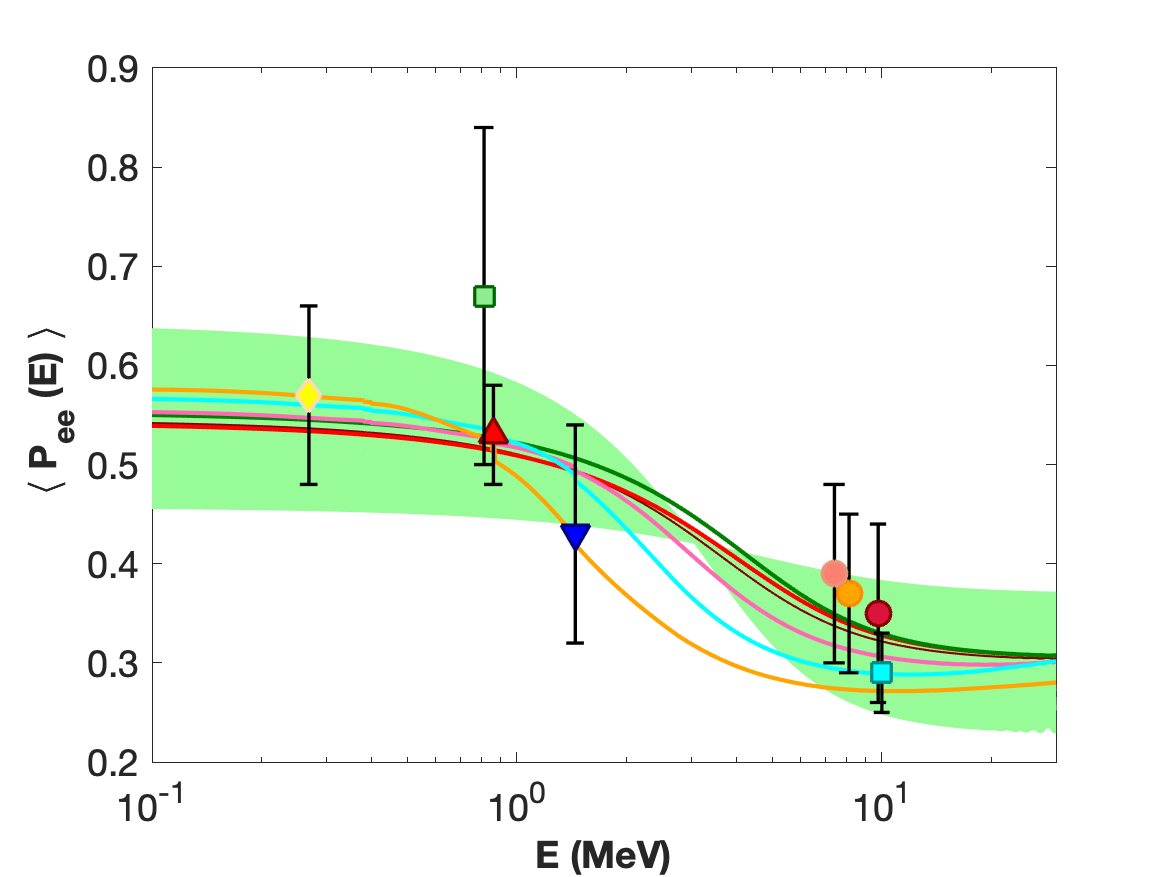}}
\caption{
The figure illustrates the averaged survival probability of $^8B$ electron neutrinos $\langle P_{ee}(E)\rangle$, resulting from the interaction of active neutrinos with the background medium. The red curve depicts the standard three-neutrino flavour oscillation model $S_{\nu}$, with a $\chi_{\nu}^2$ value of 2.73, while the blue curve represents the new neutrino-boson $R_{a\nu}$ model with $\bar{\epsilon}=10^{-3}$ and a $\chi_{\nu}^2$ value of 2.76. Both models are referenced in Table \ref{tab:neutchi2}. We also calculate a variant of the $R_{a\nu}$ model, denoted as $D_1$, where $\bar{\epsilon}=0.1$. This version is represented by a thin brown curve and a light green band with a corresponding $\chi_{\nu}^2$ value of 2.62.
Similarly, the magenta, cyan, and orange curves represent the $\langle P_{ee}(E)\rangle$ functions for models $D_2$, $D_3$, and $D_4$. These are computed at $\bar{\epsilon}=0.2$, yielding a $\chi_{\nu}^2$ value of 2.68, at $\bar{\epsilon}=0.3$ with a $\chi_{\nu}^2$ of 3.47, and at $\bar{\epsilon}=0.4$ with a $\chi_{\nu}^2$ of 4.94, respectively. All neutrino-boson models used in this figure are detailed in Table \ref{tab:neutchi2}. The solar neutrino data are consistent with the one used in Fig. \ref{fig:PeB8a}.} \label{fig:PeB8b}
\end{figure}

\medskip\noindent
To test our neutrino-boson model, we use an up-to-date standard solar model with good agreement with current neutrino fluxes and helioseismic datasets.  The details about the physics of this standard solar model in which we use the AGSS09 (low-Z) solar abundances  \citep{2009ARA&A..47..481A}   are described in \citet{2013MNRAS.435.2109L} and \citet{2020MNRAS.498.1992C}.  
The Sun's present-day structure was computed with the release version 12115 of the stellar evolution code MESA (e.g.,\citep{2019ApJS..243...10P}). 
This stellar code computes one-dimensional star structures through time; thus, the code follows the evolution of the Sun from the pre-main-sequence or zero-age main sequence until the Sun's present age, $4.57$ Gyr. Then, using a $\chi^2$— calibration optimization method, \citet{2020MNRAS.498.1992C}  obtain a present-day Sun model that better fits the observed solar values, such as the luminosity and effective temperature of the star, $ 3.8418 \times 10^{33}$ erg s$^{-1}$ and $ 5777$ K, respectively. Among other quantities, the $\chi^2$ calibration  method also fits the experimental determination of the abundance stellar surface ratio: ($Z_{\rm s}/X_{\rm s}$)$_{\odot}=0.0181$, where $Z_s$ and $X_s$ 
are the metal and hydrogen abundances at the star's surface \citep{1993ApJ...408..347T,1995RvMP...67..781B,2006ApJS..165..400B}. 

\medskip\noindent
In this nonstandard neutrino flavour oscillation model, we opt to use the parameter values corresponding to the standard three-neutrino oscillation model  \citep[e.g.,][]{2016NuPhB.908..199G}.
Hence, we adopt the recent values obtained by the data analysis 
of the standard three-neutrino flavour oscillation model obtained by
\citet{2021JHEP...02..071D}.  Accordingly, for a parameterization with  a normal ordering of neutrino masses, the mass-square difference and the mixing angles have the following values \citep[see Table 3 of][]{2021JHEP...02..071D}:	
$\Delta m^2_{21}= 7.50^{+0.22}_{-0.20}\times 10^{-5}{\rm eV^2}$, 
$\sin^2{\theta_{12}}=0.318\pm 0.016 $,
and  $\sin^2{\theta_{13}}=0.02250^{+0.00055}_{-0.00078}$.
Similarly  $\Delta m^2_{31}= 2.55^{+0.02}_{-0.03}\times 10^{-3}{\rm eV^2}$ and $\sin^2{\theta_{23}}=0.574\pm 0.014 $. Moreover,  we assume that all phases are equal to zero. 

\medskip\noindent
Figures \ref{fig:PeB8a} and  \ref{fig:PeB8b} show  the $\langle P_{ee}(E) \rangle $ [Eq. \ref{eq:Peet}] functions for different standard and nonstandard neutrino flavour oscillation models.  Table  \ref{tab:neutchi2} shows the parameters used to compute such models. 
The shape of  $\langle P_{ee}(E)\rangle $   as a function of the neutrino's energy depends on the interaction of active neutrinos with the plasma background state inside the Sun and the interaction of electron neutrinos with the boson field $\psi(t)$  inside the dark matter halo. Two parameters regulate this latter interaction:  the coupling constant $g_{\psi\nu}$ and the amplitude of the time variation boson field  $\bar{\epsilon}$. The former defines the strength of the coupling of electron neutrinos to the boson background state; the latter fixes the amplitude of the time-varying boson field on the mass differences and mixing angles of the neutrino flavour oscillation model.

\medskip\noindent
Figure \ref{fig:PeB8a} displays several models from Table \ref{tab:neutchi2} (including models $S_\nu$, $R_{a\nu}$, $R_{b\nu}$, and $A_i$ with $i=1,6$) where we choose positive and negative values of $g_{\psi\nu}$ ranging from $-7\times10^{25},G_F$ to $5\times10^{25}\,G_F$. We neglect the time variation of the boson field for now and thus set $\bar{\epsilon}=10^{-3}$. Similar to the standard three-neutrino flavour oscillation model, vacuum oscillations dominate the neutrino propagation for lower-energy neutrinos ($E\le 1\,{\rm MeV}$) inside the star, in the dark matter halo, and naturally in outer space.	
Similarly, as the neutrino energy increases ($E\ge 10\, {\rm MeV}$), the contribution from the active neutrinos' interaction with the solar background plasma (MSW effect) becomes equally important. In this study, for the nonstandard neutrino models, we include the interaction of electron neutrinos with the boson field $\psi(t)$ inside the dark matter halo. Overall, the neutrino oscillations and suppression are similar to the classical case, as shown in Fig. \ref{fig:PeB8a}. However, for those neutrino-boson models with a large value of the coupling constant $g_{\psi\nu}$, the MSW effect occurs all over the boson halo, including inside and outside the Sun, and affects the propagation of all solar neutrinos. This effect is evident in model $A_3$ (shown by the violet curve in Fig. \ref{fig:PeB8a}). Alternatively, in the case of a negative value of $g_{\psi\nu}$, the effect is reversed, as seen in model $A_6$ (shown by the brown curve in Fig. \ref{fig:PeB8a}).
 
\medskip\noindent
In addition, to highlight the impact of this new neutrino-boson model on the $\langle P_{ee}(E)\rangle $, in Figs. \ref{fig:PeB8a} and \ref{fig:PeB8b},  we show the survival probability of electron neutrinos for the standard three-neutrino flavour models (continuous red curve).  Although the standard three-neutrino flavour oscillation model agrees with the current $pp$, $pep$, $^7Be$ and $^8B$ measurements, only a restricted set of these nonstandard neutrino flavour models are in agreement with these solar neutrino measurements.  In this work, we opt to assess the quality of these new models in fitting the data by using the following $\chi_\nu^2$ statistical test:
\begin{eqnarray} 
	\chi^2_{\nu}=
	\sum^N_{i=1}
	\left(\frac{P_{ee}^{obs}(E_{i})-P_{ee}^{th}(E_{i})}{\sigma_{obs}(E_{i})} \right)^2,
	\label{eq:Chi2nu}
\end{eqnarray}  
where $N$ is the total number of data points.
The above $\chi_\nu^2$ combined the neutrino measurements made by several solar neutrino experiments at different energy values $E_i$ of the survival probability function $P_{ee}(E)$. The superscript "obs" and "th" indicate the observed and theoretical $\langle P_{ee}(E) \rangle $ [Eqs. \eqref{eq:Peet} and  \eqref{eq:Pee}] values at neutrino energy $E_{i}$, and the subscript $i$ refers to specific experimental measurement [see Fig. \eqref{fig:PeB8a}]. $\sigma_{obs}(E_{i})$  is the error of measurement $i$. The data points $P_{ee}^{obs}(E_{i})$ are measurements  obtained by
solar neutrino experiments \citep{2018Natur.562..505B,2019PhRvD.100h2004A,2010PhRvD..82c3006B,2011PhRvC..84c5804A,2016PhRvD..94e2010A,2013PhRvC..88b5501A,2008PhRvD..78c2002C}. 
For convenience, we define the degree of freedom, denoted as (d.o.f.) as ${\rm d.o.f.} = {N-k} $, where $N$ is the number of data points and $k$ is the number of parameters in the model. In this context, the reduced chi-square value, $\chi_\nu^2$, as defined in Eq. \eqref{eq:Chi2nu}, is normalized by the degree of freedom to yield $\chi_\nu^2/{\rm d.o.f.}$, effectively providing a measure of the goodness of fit per degree of freedom. In the current study, we consider $N=8$ data points. For the $S_\nu$ model, we have $k=3$ parameters, leading to a degree of freedom ${\rm d.o.f.}=5$. For all other models presented in Table \ref{tab:neutchi2}, we have $k=4$ parameters, resulting in ${\rm d.o.f.}=4$.	Figures \ref{fig:PeB8a} and \ref{fig:PeB8b} illustrate a set of data points pertinent to this analysis, while Table \ref{tab:neutchi2} lists the corresponding values of $\chi^2_{\nu}$ and $\chi^2/{\rm d.o.f.}$. 
 Interestingly, our analysis (presented in Table  \ref{tab:neutchi2}) shows that the $S_\nu$ model's $\chi^2/{\rm d.o.f.}$ value is lower than that of any other model considered in this study. Once that $S_\nu$ model contains fewer parameters, our findings suggest that the extra parameters employed in the other models may not be necessary for an accurate and robust representation of the current neutrino data. In line with the principle of Occam's razor, which advocates for simplicity in model selection, the $S_\nu$ model, with its optimal $\chi^2/{\rm d.o.f.}$  and reduced parameter count, emerges as the most favorable model.

\begin{figure} 
	$$\qquad$$	
	\centering 
	\includegraphics[scale=0.45]{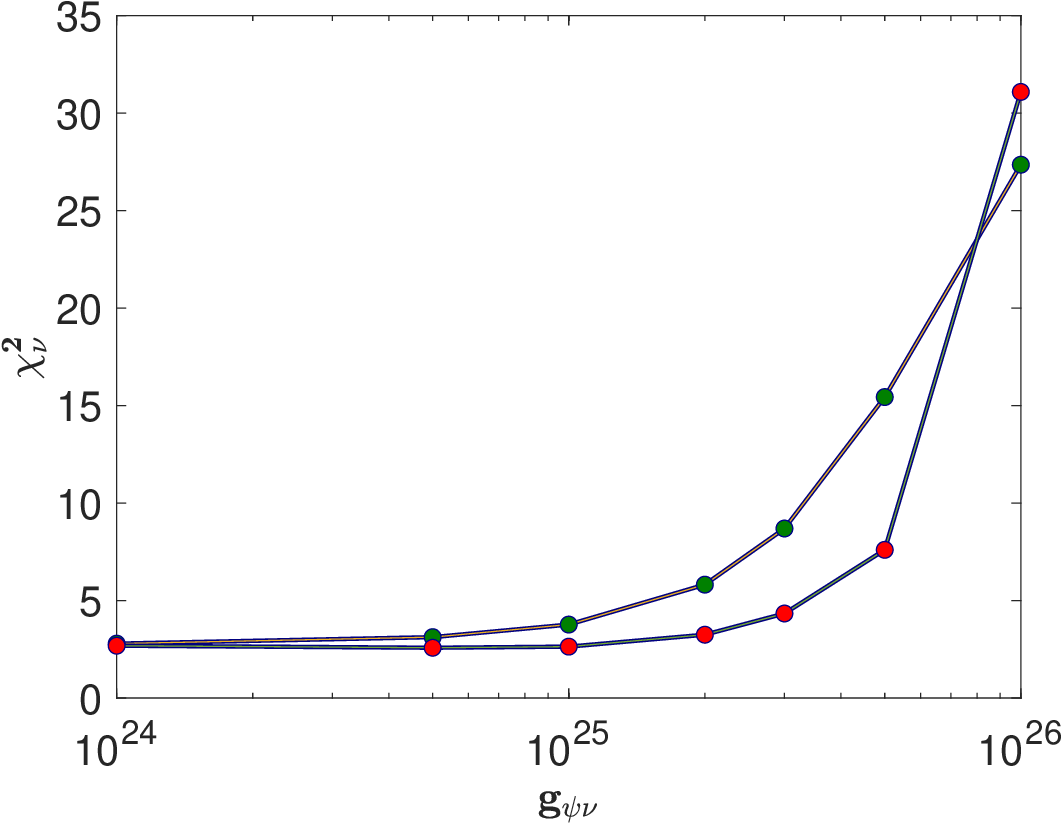}	
	\includegraphics[scale=0.45]{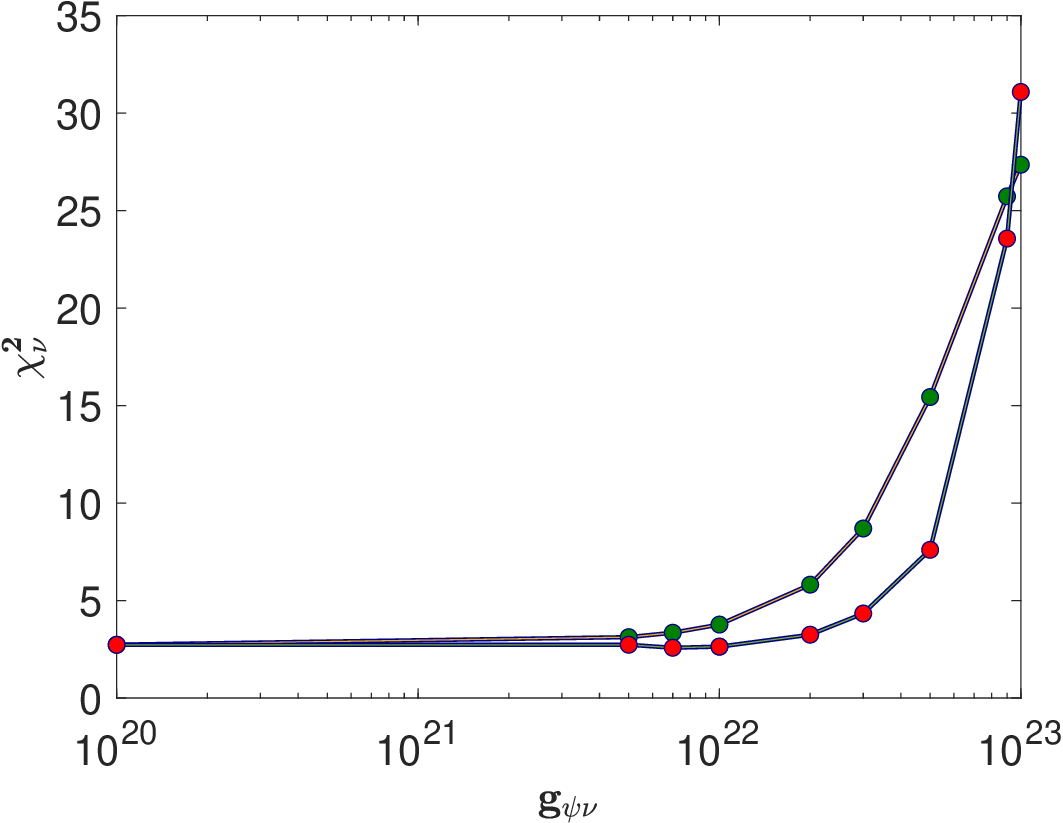}	
	\caption{The figures display the $\chi^2_\nu$ test values as a function of the coupling constant $g_{\psi \nu}$ for dark matter halos with $M_\psi = 10^{-13}\;M_\odot$ (top) and $M_\psi = 10^{-10}\;M_\odot$ (bottom), respectively. In both, green circles represent positive values of $g_{\nu}$ in the dark matter models, while red circles represent negative values.}
	\label{fig:Chi2}
\end{figure}

\medskip\noindent
However, it is noteworthy that several neutrino-boson models with specific values of $g_{\psi\nu}$ and $m_\psi$ were also found to fit the solar neutrino data, albeit not as well as the $S_\nu$ standard three-neutrino oscillation model. For instance, models $R_{b\nu}$ and $A_5$ yield $\chi_{\nu}^2$ values that are comparable to, or even greater than those of the $S_{\nu}$ model. However, these models also have larger $\chi^2/{\rm d.o.f.}$ values, indicating a less optimal fit.
Furthermore, several other models, including $A_3$, $A_4$, and $A_6$, exhibit substantially larger $\chi_{\nu}^2$ and $\chi^2/{\rm d.o.f.}$ values compared to the $S_{\nu}$ model. Given these findings, we can reasonably exclude these models from our consideration.	
Additionally, we explored the influence of $m_\psi$ and $m_\phi$ on the average survival probability, $\langle P_{ee}(E) \rangle$, for models $B_i$ and $C_i$, as indicated in the same table. However, our findings generally suggest a minimal impact. Interestingly, we identified models such as $C_1$ and $C_4$ that exhibit a $\chi_{\nu}^2$ identical to the standard model across a range of $m_\phi$ values, whether small or large.

\medskip\noindent
Figure \ref{fig:PeB8b} presents a crucial and relevant result: the interaction of electron neutrinos with the background boson field occurs through the coupling constant $g_{\psi\nu}$ [Eq. \eqref{eq:Veff}], but also depends on the amplitude $\bar{\epsilon}$ [Eq. \eqref{eq:epsilon}] of the boson field $\psi(t)$, which is determined by the time variation of mass difference and mixing angles [Eqs. \eqref{eq:deltamphi} and \eqref{eq:thetaphi}]. This figure illustrates that, compared to the standard case, the $\langle P_{ee}(E) \rangle$ of a boson-neutrino model undergoes a shift proportional to the magnitude of $\bar{\epsilon}$. This effect is visible primarily in the 1 to 10 MeV energy range. Interestingly, for some of these neutrino-boson models, $\langle P_{ee}(E) \rangle$ improves the agreement with the observational data (see models $D_i$ with $i=1,\cdots,4$ in Table \ref{tab:neutchi2}). Figure \ref{fig:PeB8b} depicts two such models, $D_1$ and $D_2$, with $\chi_{\nu}^2=2.62$ and $\chi_{\nu}^2=2.68$, respectively. These $\chi_{\nu}^2$ values are smaller than the one found for the standard case ($\chi_{\nu}^2=2.73$). Although these models are compatible with the current data, their $\chi^2/{\rm d.o.f.}$ values are larger than that of the $S_\nu$ model. Therefore, despite their compatibility, the $S_\nu$ model remains the preferred choice.

\medskip\noindent
Using $R_{a\nu}$ and $R_{b\nu}$ as reference models (see Table \ref{tab:neutchi2}), we estimate some parameter limits. Comparing models $A_i$ ($i=1,\cdots,6$) with $S_\nu$, we find that values of the neutrino-boson coupling constant $g_{\psi\nu}$ outside the interval $-10^{24}\; G_F$ and $10^{25}\; G_F$ result in $\chi_\nu^2$ values significantly higher than the $\chi_\nu^2=2.73$ one of the standard three-neutrino model. Therefore, we set the following limits for the neutrino-boson coupling constant: $-10^{24}\; G_F\le g_{\psi\nu} \le 10^{25}\; G_F$. We also find that the boson's mass $m_\psi$ can take any value in the interval $1.0\;10^{-14}$ -- $1.8\;10^{-14}\; eV$, which corresponds to neutrino-boson models $B_1$ and $B_2$ with $\chi_\nu^2=4.74$ and $\chi_\nu^2=2.80$, respectively. Finally, we conclude that the mass of the intermediate particle $m_\phi$ does not significantly affect the neutrino propagation.

\medskip\noindent
In Fig. \ref{fig:Chi2}, we plot the dependence of the parameter $g_{\psi\nu}$ on $\chi^2_\nu$ for two different dark matter halo masses: $M_\psi=10^{-10}\;M_\odot$ and $M_\psi=10^{-13}\;M_\odot$. Although $g_{\psi\nu}$ behaves similarly in both cases, we find that the optimal range of values that agrees with the solar neutrino data is different for the two dark matter halos. Specifically, for $M_\psi=10^{-10}\;M_\odot$, the agreement occurs for $g_{\psi\nu}\le 10^{22}\; G_F$, whereas for $M_\psi=10^{-13}\;M_\odot$, it occurs for $g_{\psi\nu}\le 10^{25}\;  G_F$. In both cases, we observe that the absolute value of $\chi^2_\nu$ increases rapidly as $g_{\psi\nu}$ increases. We also find that negative values of $g_{\psi\nu}$ provide a better fit to the data than positive values but for large values of $g_{\psi\nu}$, the negative solutions are overtaken by the positive ones. 
Our findings indicate that a very light dark matter halo with a mass of $M_\psi=10^{-13}\;M_\odot$ hosted by the Sun is consistent with the solar data. Furthermore, this agreement holds even for models with $g_{\psi\nu}$ negative values. However, despite the inclusion of additional parameters, the standard case $S_\nu$ remains the preferred choice based on the current solar data, as the improvement in the fitting procedure is not substantial. Additionally, we note that both negative solutions for $g_{\psi\nu}$ yield a small $\chi^2$ value, albeit with a larger $\rm \chi^2/d.o.f$. Thus, considering the results presented in Table I, it is evident that the standard model, represented as $S_\nu$, remains the most suitable option for fitting the current solar neutrino data.

\section{Summary and conclusion}
\label{sec-Summary}

\medskip\noindent
Previous studies have shown that the gravitational field of stars, including the Sun, enhances the concentration of ultralightdark matter particles in their vicinity, forming a stable dark matter halo. Our study explores explicitly the potential interaction between solar neutrinos and the ultralightdark matter particles within this halo. We investigate how the survival probability of electron neutrinos is affected when active neutrinos interact with a locally enhanced, time-dependent ultralightdark matter field, in addition to the standard interactions.
This interaction is mediated by a new particle, $\phi$. It causes active neutrinos to undergo flavour neutrino oscillations and MSW effects, as determined by the time-dependent mass term $\delta m_\nu/m_\nu$, the neutrino mixing angles, and a new term in the matter potential diagonal matrix ${\cal V}$, which defines the neutrino's interaction with the boson field $V_{\nu\phi}$.

\medskip\noindent
Our investigation reveals that the impact of solar bosonic dark matter on solar neutrinos can be categorized into two classes: (i) The effect is significant to the extent that such models can be conclusively rejected based on current data;
(ii) The impact is minimal, and the additional freedom does not improve the agreement between the models and the data.

\medskip\noindent
Here, we showcase the robustness of the current agreement between the standard flavour oscillation model and the data, making it very difficult to challenge. Even with the incorporation of ultralightbosons, substantial couplings, and the inclusion of a mediator, the standard model maintains its strong compatibility with the data. This finding aligns with previous research by  \citep{2020ApJ...905...22L}, which also revealed a similar outcome regarding the average local dark matter density. Our study further confirms that this challenge persists even when considering a highly dense halo confined within the Sun. However, it is worth noting that as more data become available, it remains possible to disregard these models or discover potential improvements in the future.
The validation of such a class of models will be tested by the next generation of neutrino detectors, including hybrid optical neutrino detectors such as 
{\sc Theia} \citep{2020EPJC...80..416A}, Jinping \citep{2017ARNPS..67..231C}
and {\sc Yemilab} \citep{2019arXiv190305368S}, or by liquid scintillator experiments like {\sc Juno} \citep{2016JPhG...43c0401A,2022PrPNP.12303927J} and {\sc SNO+} 
\citep{2020arXiv201112924S}, and Water Cherenkov experiments such as
Hyper-Kamiokande \citep{2018arXiv180504163H} and {\sc Dune}
\citep{2016arXiv160102984A}.

\medskip\noindent

\begin{acknowledgments}
I.L. thanks the Funda\c c\~ao para a Ci\^encia e Tecnologia (FCT), Portugal, for the financial support to the Center for Astrophysics and Gravitation (CENTRA/IST/ULisboa)  through the Grant Project~No.~UIDB/00099/2020  and Grant No. PTDC/FIS-AST/28920/2017.  
\end{acknowledgments}


\begin{thebibliography}{105}
	\expandafter\ifx\csname natexlab\endcsname\relax\def\natexlab#1{#1}\fi
	\expandafter\ifx\csname bibnamefont\endcsname\relax
	\def\bibnamefont#1{#1}\fi
	\expandafter\ifx\csname bibfnamefont\endcsname\relax
	\def\bibfnamefont#1{#1}\fi
	\expandafter\ifx\csname citenamefont\endcsname\relax
	\def\citenamefont#1{#1}\fi
	\expandafter\ifx\csname url\endcsname\relax
	\def\url#1{\texttt{#1}}\fi
	\expandafter\ifx\csname urlprefix\endcsname\relax\def\urlprefix{URL }\fi
	\providecommand{\bibinfo}[2]{#2}
	\providecommand{\eprint}[2][]{\url{#2}}
	
	\bibitem[{\citenamefont{{Zwicky}}(1933)}]{1933AcHPh...6..110Z}
	\bibinfo{author}{\bibfnamefont{F.}~\bibnamefont{{Zwicky}}},
	\bibinfo{journal}{Helvetica Physica Acta} \textbf{\bibinfo{volume}{6}},
	\bibinfo{pages}{110} (\bibinfo{year}{1933}).
	
	\bibitem[{\citenamefont{{Battaglieri} et~al.}(2017)\citenamefont{{Battaglieri},
			{Belloni}, {Chou}, {Cushman}, {Echenard}, {Essig}, {Estrada}, {Feng},
			{Flaugher}, {Fox} et~al.}}]{2017arXiv170704591B}
	\bibinfo{author}{\bibfnamefont{M.}~\bibnamefont{{Battaglieri}}},
	\bibinfo{author}{\bibfnamefont{A.}~\bibnamefont{{Belloni}}},
	\bibinfo{author}{\bibfnamefont{A.}~\bibnamefont{{Chou}}},
	\bibinfo{author}{\bibfnamefont{P.}~\bibnamefont{{Cushman}}},
	\bibinfo{author}{\bibfnamefont{B.}~\bibnamefont{{Echenard}}},
	\bibinfo{author}{\bibfnamefont{R.}~\bibnamefont{{Essig}}},
	\bibinfo{author}{\bibfnamefont{J.}~\bibnamefont{{Estrada}}},
	\bibinfo{author}{\bibfnamefont{J.~L.} \bibnamefont{{Feng}}},
	\bibinfo{author}{\bibfnamefont{B.}~\bibnamefont{{Flaugher}}},
	\bibinfo{author}{\bibfnamefont{P.~J.} \bibnamefont{{Fox}}},
	\bibnamefont{et~al.}, \bibinfo{journal}{arXiv e-prints}
	\bibinfo{eid}{arXiv:1707.04591} (\bibinfo{year}{2017}), \eprint{1707.04591}.
	
	\bibitem[{\citenamefont{{Feng}}(2010)}]{2010ARA&A..48..495F}
	\bibinfo{author}{\bibfnamefont{J.~L.} \bibnamefont{{Feng}}},
	\bibinfo{journal}{\araa} \textbf{\bibinfo{volume}{48}}, \bibinfo{pages}{495}
	(\bibinfo{year}{2010}), \eprint{1003.0904}.
	
	\bibitem[{\citenamefont{{Baer} et~al.}(2015)\citenamefont{{Baer}, {Choi},
			{Kim}, and {Roszkowski}}}]{2015PhR...555....1B}
	\bibinfo{author}{\bibfnamefont{H.}~\bibnamefont{{Baer}}},
	\bibinfo{author}{\bibfnamefont{K.-Y.} \bibnamefont{{Choi}}},
	\bibinfo{author}{\bibfnamefont{J.~E.} \bibnamefont{{Kim}}}, \bibnamefont{and}
	\bibinfo{author}{\bibfnamefont{L.}~\bibnamefont{{Roszkowski}}},
	\bibinfo{journal}{Phys.~Rep.} \textbf{\bibinfo{volume}{555}},
	\bibinfo{pages}{1} (\bibinfo{year}{2015}), \eprint{1407.0017}.
	
	\bibitem[{\citenamefont{{Agrawal} et~al.}(2021)\citenamefont{{Agrawal},
			{Bauer}, {Beacham}, {Berlin}, {Boyarsky}, {Cebrian}, {Cid-Vidal},
			{d'Enterria}, {De Roeck}, {Drewes} et~al.}}]{2021EPJC...81.1015A}
	\bibinfo{author}{\bibfnamefont{P.}~\bibnamefont{{Agrawal}}},
	\bibinfo{author}{\bibfnamefont{M.}~\bibnamefont{{Bauer}}},
	\bibinfo{author}{\bibfnamefont{J.}~\bibnamefont{{Beacham}}},
	\bibinfo{author}{\bibfnamefont{A.}~\bibnamefont{{Berlin}}},
	\bibinfo{author}{\bibfnamefont{A.}~\bibnamefont{{Boyarsky}}},
	\bibinfo{author}{\bibfnamefont{S.}~\bibnamefont{{Cebrian}}},
	\bibinfo{author}{\bibfnamefont{X.}~\bibnamefont{{Cid-Vidal}}},
	\bibinfo{author}{\bibfnamefont{D.}~\bibnamefont{{d'Enterria}}},
	\bibinfo{author}{\bibfnamefont{A.}~\bibnamefont{{De Roeck}}},
	\bibinfo{author}{\bibfnamefont{M.}~\bibnamefont{{Drewes}}},
	\bibnamefont{et~al.}, \bibinfo{journal}{European Physical Journal C}
	\textbf{\bibinfo{volume}{81}}, \bibinfo{eid}{1015} (\bibinfo{year}{2021}),
	\eprint{2102.12143}.
	
	\bibitem[{\citenamefont{{Graham} et~al.}(2015)\citenamefont{{Graham},
			{Irastorza}, {Lamoreaux}, {Lindner}, and {van Bibber}}}]{2015ARNPS..65..485G}
	\bibinfo{author}{\bibfnamefont{P.~W.} \bibnamefont{{Graham}}},
	\bibinfo{author}{\bibfnamefont{I.~G.} \bibnamefont{{Irastorza}}},
	\bibinfo{author}{\bibfnamefont{S.~K.} \bibnamefont{{Lamoreaux}}},
	\bibinfo{author}{\bibfnamefont{A.}~\bibnamefont{{Lindner}}},
	\bibnamefont{and} \bibinfo{author}{\bibfnamefont{K.~A.} \bibnamefont{{van
				Bibber}}}, \bibinfo{journal}{Annual Review of Nuclear and Particle Science}
	\textbf{\bibinfo{volume}{65}}, \bibinfo{pages}{485} (\bibinfo{year}{2015}),
	\eprint{1602.00039}.
	
	\bibitem[{\citenamefont{{Ackerman} et~al.}(2009)\citenamefont{{Ackerman},
			{Buckley}, {Carroll}, and {Kamionkowski}}}]{2009PhRvD..79b3519A}
	\bibinfo{author}{\bibfnamefont{L.}~\bibnamefont{{Ackerman}}},
	\bibinfo{author}{\bibfnamefont{M.~R.} \bibnamefont{{Buckley}}},
	\bibinfo{author}{\bibfnamefont{S.~M.} \bibnamefont{{Carroll}}},
	\bibnamefont{and}
	\bibinfo{author}{\bibfnamefont{M.}~\bibnamefont{{Kamionkowski}}},
	\bibinfo{journal}{\prd} \textbf{\bibinfo{volume}{79}}, \bibinfo{eid}{023519}
	(\bibinfo{year}{2009}), \eprint{0810.5126}.
	
	\bibitem[{\citenamefont{{Di Luzio} et~al.}(2020)\citenamefont{{Di Luzio},
			{Giannotti}, {Nardi}, and {Visinelli}}}]{2020PhR...870....1D}
	\bibinfo{author}{\bibfnamefont{L.}~\bibnamefont{{Di Luzio}}},
	\bibinfo{author}{\bibfnamefont{M.}~\bibnamefont{{Giannotti}}},
	\bibinfo{author}{\bibfnamefont{E.}~\bibnamefont{{Nardi}}}, \bibnamefont{and}
	\bibinfo{author}{\bibfnamefont{L.}~\bibnamefont{{Visinelli}}},
	\bibinfo{journal}{Physics Reports} \textbf{\bibinfo{volume}{870}},
	\bibinfo{pages}{1} (\bibinfo{year}{2020}), \eprint{2003.01100}.
	
	\bibitem[{\citenamefont{{Marsh}}(2016)}]{2016PhR...643....1M}
	\bibinfo{author}{\bibfnamefont{D.~J.~E.} \bibnamefont{{Marsh}}},
	\bibinfo{journal}{Physics Reports} \textbf{\bibinfo{volume}{643}},
	\bibinfo{pages}{1} (\bibinfo{year}{2016}), \eprint{1510.07633}.
	
	\bibitem[{\citenamefont{{Asztalos} et~al.}(2010)\citenamefont{{Asztalos},
			{Carosi}, {Hagmann}, {Kinion}, {van Bibber}, {Hotz}, {Rosenberg}, {Rybka},
			{Hoskins}, {Hwang} et~al.}}]{2010PhRvL.104d1301A}
	\bibinfo{author}{\bibfnamefont{S.~J.} \bibnamefont{{Asztalos}}},
	\bibinfo{author}{\bibfnamefont{G.}~\bibnamefont{{Carosi}}},
	\bibinfo{author}{\bibfnamefont{C.}~\bibnamefont{{Hagmann}}},
	\bibinfo{author}{\bibfnamefont{D.}~\bibnamefont{{Kinion}}},
	\bibinfo{author}{\bibfnamefont{K.}~\bibnamefont{{van Bibber}}},
	\bibinfo{author}{\bibfnamefont{M.}~\bibnamefont{{Hotz}}},
	\bibinfo{author}{\bibfnamefont{L.~J.} \bibnamefont{{Rosenberg}}},
	\bibinfo{author}{\bibfnamefont{G.}~\bibnamefont{{Rybka}}},
	\bibinfo{author}{\bibfnamefont{J.}~\bibnamefont{{Hoskins}}},
	\bibinfo{author}{\bibfnamefont{J.}~\bibnamefont{{Hwang}}},
	\bibnamefont{et~al.}, \bibinfo{journal}{\prl} \textbf{\bibinfo{volume}{104}},
	\bibinfo{eid}{041301} (\bibinfo{year}{2010}), \eprint{0910.5914}.
	
	\bibitem[{\citenamefont{{Budker} et~al.}(2014)\citenamefont{{Budker}, {Graham},
			{Ledbetter}, {Rajendran}, and {Sushkov}}}]{2014PhRvX...4b1030B}
	\bibinfo{author}{\bibfnamefont{D.}~\bibnamefont{{Budker}}},
	\bibinfo{author}{\bibfnamefont{P.~W.} \bibnamefont{{Graham}}},
	\bibinfo{author}{\bibfnamefont{M.}~\bibnamefont{{Ledbetter}}},
	\bibinfo{author}{\bibfnamefont{S.}~\bibnamefont{{Rajendran}}},
	\bibnamefont{and} \bibinfo{author}{\bibfnamefont{A.~O.}
		\bibnamefont{{Sushkov}}}, \bibinfo{journal}{Physical Review X}
	\textbf{\bibinfo{volume}{4}}, \bibinfo{eid}{021030} (\bibinfo{year}{2014}),
	\eprint{1306.6089}.
	
	\bibitem[{\citenamefont{{Stadnik} and {Flambaum}}(2014)}]{2014PhRvD..89d3522S}
	\bibinfo{author}{\bibfnamefont{Y.~V.} \bibnamefont{{Stadnik}}}
	\bibnamefont{and} \bibinfo{author}{\bibfnamefont{V.~V.}
		\bibnamefont{{Flambaum}}}, \bibinfo{journal}{\prd}
	\textbf{\bibinfo{volume}{89}}, \bibinfo{eid}{043522} (\bibinfo{year}{2014}),
	\eprint{1312.6667}.
	
	\bibitem[{\citenamefont{{Orebi Gann} et~al.}(2021)\citenamefont{{Orebi Gann},
			{Zuber}, {Bemmerer}, and {Serenelli}}}]{2021ARNPS..71..491O}
	\bibinfo{author}{\bibfnamefont{G.~D.} \bibnamefont{{Orebi Gann}}},
	\bibinfo{author}{\bibfnamefont{K.}~\bibnamefont{{Zuber}}},
	\bibinfo{author}{\bibfnamefont{D.}~\bibnamefont{{Bemmerer}}},
	\bibnamefont{and}
	\bibinfo{author}{\bibfnamefont{A.}~\bibnamefont{{Serenelli}}},
	\bibinfo{journal}{Annual Review of Nuclear and Particle Science}
	\textbf{\bibinfo{volume}{71}}, \bibinfo{pages}{491} (\bibinfo{year}{2021}),
	\eprint{2107.08613}.
	
	\bibitem[{\citenamefont{{Dror} et~al.}(2021)\citenamefont{{Dror}, {Murayama},
			and {Rodd}}}]{2021PhRvD.103k5004D}
	\bibinfo{author}{\bibfnamefont{J.~A.} \bibnamefont{{Dror}}},
	\bibinfo{author}{\bibfnamefont{H.}~\bibnamefont{{Murayama}}},
	\bibnamefont{and} \bibinfo{author}{\bibfnamefont{N.~L.}
		\bibnamefont{{Rodd}}}, \bibinfo{journal}{\prd}
	\textbf{\bibinfo{volume}{103}}, \bibinfo{eid}{115004} (\bibinfo{year}{2021}),
	\eprint{2101.09287}.
	
	\bibitem[{\citenamefont{{Billard} et~al.}(2022)\citenamefont{{Billard},
			{Boulay}, {Cebri{\'a}n}, {Covi}, {Fiorillo}, {Green}, {Kopp}, {Majorovits},
			{Palladino}, {Petricca} et~al.}}]{2022RPPh...85e6201B}
	\bibinfo{author}{\bibfnamefont{J.}~\bibnamefont{{Billard}}},
	\bibinfo{author}{\bibfnamefont{M.}~\bibnamefont{{Boulay}}},
	\bibinfo{author}{\bibfnamefont{S.}~\bibnamefont{{Cebri{\'a}n}}},
	\bibinfo{author}{\bibfnamefont{L.}~\bibnamefont{{Covi}}},
	\bibinfo{author}{\bibfnamefont{G.}~\bibnamefont{{Fiorillo}}},
	\bibinfo{author}{\bibfnamefont{A.}~\bibnamefont{{Green}}},
	\bibinfo{author}{\bibfnamefont{J.}~\bibnamefont{{Kopp}}},
	\bibinfo{author}{\bibfnamefont{B.}~\bibnamefont{{Majorovits}}},
	\bibinfo{author}{\bibfnamefont{K.}~\bibnamefont{{Palladino}}},
	\bibinfo{author}{\bibfnamefont{F.}~\bibnamefont{{Petricca}}},
	\bibnamefont{et~al.}, \bibinfo{journal}{Reports on Progress in Physics}
	\textbf{\bibinfo{volume}{85}}, \bibinfo{eid}{056201} (\bibinfo{year}{2022}),
	\eprint{2104.07634}.
	
	\bibitem[{\citenamefont{{Hui} et~al.}(2017)\citenamefont{{Hui}, {Ostriker},
			{Tremaine}, and {Witten}}}]{2017PhRvD..95d3541H}
	\bibinfo{author}{\bibfnamefont{L.}~\bibnamefont{{Hui}}},
	\bibinfo{author}{\bibfnamefont{J.~P.} \bibnamefont{{Ostriker}}},
	\bibinfo{author}{\bibfnamefont{S.}~\bibnamefont{{Tremaine}}},
	\bibnamefont{and} \bibinfo{author}{\bibfnamefont{E.}~\bibnamefont{{Witten}}},
	\bibinfo{journal}{\prd} \textbf{\bibinfo{volume}{95}}, \bibinfo{eid}{043541}
	(\bibinfo{year}{2017}), \eprint{1610.08297}.
	
	\bibitem[{\citenamefont{{Niemeyer}}(2020)}]{2020PrPNP.11303787N}
	\bibinfo{author}{\bibfnamefont{J.~C.} \bibnamefont{{Niemeyer}}},
	\bibinfo{journal}{Progress in Particle and Nuclear Physics}
	\textbf{\bibinfo{volume}{113}}, \bibinfo{eid}{103787} (\bibinfo{year}{2020}),
	\eprint{1912.07064}.
	
	\bibitem[{\citenamefont{{Catena} and {Ullio}}(2010)}]{2010JCAP...08..004C}
	\bibinfo{author}{\bibfnamefont{R.}~\bibnamefont{{Catena}}} \bibnamefont{and}
	\bibinfo{author}{\bibfnamefont{P.}~\bibnamefont{{Ullio}}},
	\bibinfo{journal}{\jcap} \textbf{\bibinfo{volume}{2010}}, \bibinfo{eid}{004}
	(\bibinfo{year}{2010}), \eprint{0907.0018}.
	
	\bibitem[{\citenamefont{{Planck Collaboration}
			et~al.}(2020)\citenamefont{{Planck Collaboration}, {Aghanim}, {Akrami},
			{Ashdown}, {Aumont}, {Baccigalupi}, {Ballardini}, {Banday}, {Barreiro},
			{Bartolo} et~al.}}]{2020A&A...641A...6P}
	\bibinfo{author}{\bibnamefont{{Planck Collaboration}}},
	\bibinfo{author}{\bibfnamefont{N.}~\bibnamefont{{Aghanim}}},
	\bibinfo{author}{\bibfnamefont{Y.}~\bibnamefont{{Akrami}}},
	\bibinfo{author}{\bibfnamefont{M.}~\bibnamefont{{Ashdown}}},
	\bibinfo{author}{\bibfnamefont{J.}~\bibnamefont{{Aumont}}},
	\bibinfo{author}{\bibfnamefont{C.}~\bibnamefont{{Baccigalupi}}},
	\bibinfo{author}{\bibfnamefont{M.}~\bibnamefont{{Ballardini}}},
	\bibinfo{author}{\bibfnamefont{A.~J.} \bibnamefont{{Banday}}},
	\bibinfo{author}{\bibfnamefont{R.~B.} \bibnamefont{{Barreiro}}},
	\bibinfo{author}{\bibfnamefont{N.}~\bibnamefont{{Bartolo}}},
	\bibnamefont{et~al.}, \bibinfo{journal}{\aap} \textbf{\bibinfo{volume}{641}},
	\bibinfo{eid}{A6} (\bibinfo{year}{2020}), \eprint{1807.06209}.
	
	\bibitem[{\citenamefont{{Lopes} and
			{Turck-Chi{\`e}ze}}(2013)}]{2013ApJ...765...14L}
	\bibinfo{author}{\bibfnamefont{I.}~\bibnamefont{{Lopes}}} \bibnamefont{and}
	\bibinfo{author}{\bibfnamefont{S.}~\bibnamefont{{Turck-Chi{\`e}ze}}},
	\bibinfo{journal}{\apj} \textbf{\bibinfo{volume}{765}}, \bibinfo{eid}{14}
	(\bibinfo{year}{2013}), \eprint{1302.2791}.
	
	\bibitem[{\citenamefont{{Abbott} and {Sikivie}}(1983)}]{1983PhLB..120..133A}
	\bibinfo{author}{\bibfnamefont{L.~F.} \bibnamefont{{Abbott}}} \bibnamefont{and}
	\bibinfo{author}{\bibfnamefont{P.}~\bibnamefont{{Sikivie}}},
	\bibinfo{journal}{Physics Letters B} \textbf{\bibinfo{volume}{120}},
	\bibinfo{pages}{133} (\bibinfo{year}{1983}).
	
	\bibitem[{\citenamefont{{Khmelnitsky} and
			{Rubakov}}(2014)}]{2014JCAP...02..019K}
	\bibinfo{author}{\bibfnamefont{A.}~\bibnamefont{{Khmelnitsky}}}
	\bibnamefont{and}
	\bibinfo{author}{\bibfnamefont{V.}~\bibnamefont{{Rubakov}}},
	\bibinfo{journal}{\jcap} \textbf{\bibinfo{volume}{2014}}, \bibinfo{eid}{019}
	(\bibinfo{year}{2014}), \eprint{1309.5888}.
	
	\bibitem[{\citenamefont{{Blas} et~al.}(2017)\citenamefont{{Blas}, {Nacir}, and
			{Sibiryakov}}}]{2017PhRvL.118z1102B}
	\bibinfo{author}{\bibfnamefont{D.}~\bibnamefont{{Blas}}},
	\bibinfo{author}{\bibfnamefont{D.~L.} \bibnamefont{{Nacir}}},
	\bibnamefont{and}
	\bibinfo{author}{\bibfnamefont{S.}~\bibnamefont{{Sibiryakov}}},
	\bibinfo{journal}{\prl} \textbf{\bibinfo{volume}{118}}, \bibinfo{eid}{261102}
	(\bibinfo{year}{2017}), \eprint{1612.06789}.
	
	\bibitem[{\citenamefont{{Colpi} et~al.}(1986)\citenamefont{{Colpi}, {Shapiro},
			and {Wasserman}}}]{1986PhRvL..57.2485C}
	\bibinfo{author}{\bibfnamefont{M.}~\bibnamefont{{Colpi}}},
	\bibinfo{author}{\bibfnamefont{S.~L.} \bibnamefont{{Shapiro}}},
	\bibnamefont{and}
	\bibinfo{author}{\bibfnamefont{I.}~\bibnamefont{{Wasserman}}},
	\bibinfo{journal}{\prl} \textbf{\bibinfo{volume}{57}}, \bibinfo{pages}{2485}
	(\bibinfo{year}{1986}).
	
	\bibitem[{\citenamefont{{Liebling} and
			{Palenzuela}}(2012)}]{2012LRR....15....6L}
	\bibinfo{author}{\bibfnamefont{S.~L.} \bibnamefont{{Liebling}}}
	\bibnamefont{and}
	\bibinfo{author}{\bibfnamefont{C.}~\bibnamefont{{Palenzuela}}},
	\bibinfo{journal}{Living Reviews in Relativity}
	\textbf{\bibinfo{volume}{15}}, \bibinfo{eid}{6} (\bibinfo{year}{2012}),
	\eprint{1202.5809}.
	
	\bibitem[{\citenamefont{{Kaup}}(1968)}]{1968PhRv..172.1331K}
	\bibinfo{author}{\bibfnamefont{D.~J.} \bibnamefont{{Kaup}}},
	\bibinfo{journal}{Physical Review} \textbf{\bibinfo{volume}{172}},
	\bibinfo{pages}{1331} (\bibinfo{year}{1968}).
	
	\bibitem[{\citenamefont{{Ruffini} and {Bonazzola}}(1969)}]{1969PhRv..187.1767R}
	\bibinfo{author}{\bibfnamefont{R.}~\bibnamefont{{Ruffini}}} \bibnamefont{and}
	\bibinfo{author}{\bibfnamefont{S.}~\bibnamefont{{Bonazzola}}},
	\bibinfo{journal}{Physical Review} \textbf{\bibinfo{volume}{187}},
	\bibinfo{pages}{1767} (\bibinfo{year}{1969}).
	
	\bibitem[{\citenamefont{{Wheeler}}(1955)}]{1955PhRv...97..511W}
	\bibinfo{author}{\bibfnamefont{J.~A.} \bibnamefont{{Wheeler}}},
	\bibinfo{journal}{Physical Review} \textbf{\bibinfo{volume}{97}},
	\bibinfo{pages}{511} (\bibinfo{year}{1955}).
	
	\bibitem[{\citenamefont{{Braaten} and {Zhang}}(2019)}]{2019RvMP...91d1002B}
	\bibinfo{author}{\bibfnamefont{E.}~\bibnamefont{{Braaten}}} \bibnamefont{and}
	\bibinfo{author}{\bibfnamefont{H.}~\bibnamefont{{Zhang}}},
	\bibinfo{journal}{Reviews of Modern Physics} \textbf{\bibinfo{volume}{91}},
	\bibinfo{eid}{041002} (\bibinfo{year}{2019}).
	
	\bibitem[{\citenamefont{{Namjoo} et~al.}(2018)\citenamefont{{Namjoo}, {Guth},
			and {Kaiser}}}]{2018PhRvD..98a6011N}
	\bibinfo{author}{\bibfnamefont{M.~H.} \bibnamefont{{Namjoo}}},
	\bibinfo{author}{\bibfnamefont{A.~H.} \bibnamefont{{Guth}}},
	\bibnamefont{and} \bibinfo{author}{\bibfnamefont{D.~I.}
		\bibnamefont{{Kaiser}}}, \bibinfo{journal}{\prd}
	\textbf{\bibinfo{volume}{98}}, \bibinfo{eid}{016011} (\bibinfo{year}{2018}),
	\eprint{1712.00445}.
	
	\bibitem[{\citenamefont{{Chavanis}}(2011)}]{2011PhRvD..84d3531C}
	\bibinfo{author}{\bibfnamefont{P.-H.} \bibnamefont{{Chavanis}}},
	\bibinfo{journal}{\prd} \textbf{\bibinfo{volume}{84}}, \bibinfo{eid}{043531}
	(\bibinfo{year}{2011}), \eprint{1103.2050}.
	
	\bibitem[{\citenamefont{{Eby} et~al.}(2018)\citenamefont{{Eby}, {Leembruggen},
			{Street}, {Suranyi}, and {Wijewardhana}}}]{2018PhRvD..98l3013E}
	\bibinfo{author}{\bibfnamefont{J.}~\bibnamefont{{Eby}}},
	\bibinfo{author}{\bibfnamefont{M.}~\bibnamefont{{Leembruggen}}},
	\bibinfo{author}{\bibfnamefont{L.}~\bibnamefont{{Street}}},
	\bibinfo{author}{\bibfnamefont{P.}~\bibnamefont{{Suranyi}}},
	\bibnamefont{and} \bibinfo{author}{\bibfnamefont{L.~C.~R.}
		\bibnamefont{{Wijewardhana}}}, \bibinfo{journal}{\prd}
	\textbf{\bibinfo{volume}{98}}, \bibinfo{eid}{123013} (\bibinfo{year}{2018}),
	\eprint{1809.08598}.
	
	\bibitem[{\citenamefont{{B{\"o}hmer} and {Harko}}(2007)}]{2007JCAP...06..025B}
	\bibinfo{author}{\bibfnamefont{C.~G.} \bibnamefont{{B{\"o}hmer}}}
	\bibnamefont{and} \bibinfo{author}{\bibfnamefont{T.}~\bibnamefont{{Harko}}},
	\bibinfo{journal}{\jcap} \textbf{\bibinfo{volume}{2007}}, \bibinfo{eid}{025}
	(\bibinfo{year}{2007}), \eprint{0705.4158}.
	
	\bibitem[{\citenamefont{{Banerjee}
			et~al.}(2020{\natexlab{a}})\citenamefont{{Banerjee}, {Budker}, {Eby},
			{Flambaum}, {Kim}, {Matsedonskyi}, and {Perez}}}]{2020JHEP...09..004B}
	\bibinfo{author}{\bibfnamefont{A.}~\bibnamefont{{Banerjee}}},
	\bibinfo{author}{\bibfnamefont{D.}~\bibnamefont{{Budker}}},
	\bibinfo{author}{\bibfnamefont{J.}~\bibnamefont{{Eby}}},
	\bibinfo{author}{\bibfnamefont{V.~V.} \bibnamefont{{Flambaum}}},
	\bibinfo{author}{\bibfnamefont{H.}~\bibnamefont{{Kim}}},
	\bibinfo{author}{\bibfnamefont{O.}~\bibnamefont{{Matsedonskyi}}},
	\bibnamefont{and} \bibinfo{author}{\bibfnamefont{G.}~\bibnamefont{{Perez}}},
	\bibinfo{journal}{Journal of High Energy Physics}
	\textbf{\bibinfo{volume}{2020}}, \bibinfo{eid}{4}
	(\bibinfo{year}{2020}{\natexlab{a}}), \eprint{1912.04295}.
	
	\bibitem[{\citenamefont{{Banerjee}
			et~al.}(2020{\natexlab{b}})\citenamefont{{Banerjee}, {Budker}, {Eby}, {Kim},
			and {Perez}}}]{2020CmPhy...3....1B}
	\bibinfo{author}{\bibfnamefont{A.}~\bibnamefont{{Banerjee}}},
	\bibinfo{author}{\bibfnamefont{D.}~\bibnamefont{{Budker}}},
	\bibinfo{author}{\bibfnamefont{J.}~\bibnamefont{{Eby}}},
	\bibinfo{author}{\bibfnamefont{H.}~\bibnamefont{{Kim}}}, \bibnamefont{and}
	\bibinfo{author}{\bibfnamefont{G.}~\bibnamefont{{Perez}}},
	\bibinfo{journal}{Communications Physics} \textbf{\bibinfo{volume}{3}},
	\bibinfo{eid}{1} (\bibinfo{year}{2020}{\natexlab{b}}), \eprint{1902.08212}.
	
	\bibitem[{\citenamefont{{Lopes} and {Silk}}(2013)}]{2013MNRAS.435.2109L}
	\bibinfo{author}{\bibfnamefont{I.}~\bibnamefont{{Lopes}}} \bibnamefont{and}
	\bibinfo{author}{\bibfnamefont{J.}~\bibnamefont{{Silk}}},
	\bibinfo{journal}{\mnras} \textbf{\bibinfo{volume}{435}},
	\bibinfo{pages}{2109} (\bibinfo{year}{2013}), \eprint{1309.7571}.
	
	\bibitem[{\citenamefont{{Miranda} et~al.}(2015)\citenamefont{{Miranda},
			{Moura}, and {Parada}}}]{2015PhLB..744...55M}
	\bibinfo{author}{\bibfnamefont{O.~G.} \bibnamefont{{Miranda}}},
	\bibinfo{author}{\bibfnamefont{C.~A.} \bibnamefont{{Moura}}},
	\bibnamefont{and} \bibinfo{author}{\bibfnamefont{A.}~\bibnamefont{{Parada}}},
	\bibinfo{journal}{Physics Letters B} \textbf{\bibinfo{volume}{744}},
	\bibinfo{pages}{55} (\bibinfo{year}{2015}).
	
	\bibitem[{\citenamefont{{Arg{\"u}elles}
			et~al.}(2020)\citenamefont{{Arg{\"u}elles}, {Aurisano}, {Batell}, {Berger},
			{Bishai}, {Boschi}, {Byrnes}, {Chatterjee}, {Chodos}, {Coan}
			et~al.}}]{2020RPPh...83l4201A}
	\bibinfo{author}{\bibfnamefont{C.~A.} \bibnamefont{{Arg{\"u}elles}}},
	\bibinfo{author}{\bibfnamefont{A.~J.} \bibnamefont{{Aurisano}}},
	\bibinfo{author}{\bibfnamefont{B.}~\bibnamefont{{Batell}}},
	\bibinfo{author}{\bibfnamefont{J.}~\bibnamefont{{Berger}}},
	\bibinfo{author}{\bibfnamefont{M.}~\bibnamefont{{Bishai}}},
	\bibinfo{author}{\bibfnamefont{T.}~\bibnamefont{{Boschi}}},
	\bibinfo{author}{\bibfnamefont{N.}~\bibnamefont{{Byrnes}}},
	\bibinfo{author}{\bibfnamefont{A.}~\bibnamefont{{Chatterjee}}},
	\bibinfo{author}{\bibfnamefont{A.}~\bibnamefont{{Chodos}}},
	\bibinfo{author}{\bibfnamefont{T.}~\bibnamefont{{Coan}}},
	\bibnamefont{et~al.}, \bibinfo{journal}{Reports on Progress in Physics}
	\textbf{\bibinfo{volume}{83}}, \bibinfo{eid}{124201} (\bibinfo{year}{2020}),
	\eprint{1907.08311}.
	
	\bibitem[{\citenamefont{{Gonzalez-Garcia} and
			{Maltoni}}(2008)}]{2008PhR...460....1G}
	\bibinfo{author}{\bibfnamefont{M.~C.} \bibnamefont{{Gonzalez-Garcia}}}
	\bibnamefont{and}
	\bibinfo{author}{\bibfnamefont{M.}~\bibnamefont{{Maltoni}}},
	\bibinfo{journal}{Physics Reports} \textbf{\bibinfo{volume}{460}},
	\bibinfo{pages}{1} (\bibinfo{year}{2008}), \eprint{0704.1800}.
	
	\bibitem[{\citenamefont{{Smirnov} and {Xu}}(2019)}]{2019JHEP...12..046S}
	\bibinfo{author}{\bibfnamefont{A.~Y.} \bibnamefont{{Smirnov}}}
	\bibnamefont{and} \bibinfo{author}{\bibfnamefont{X.-J.} \bibnamefont{{Xu}}},
	\bibinfo{journal}{Journal of High Energy Physics}
	\textbf{\bibinfo{volume}{2019}}, \bibinfo{eid}{46} (\bibinfo{year}{2019}),
	\eprint{1909.07505}.
	
	\bibitem[{\citenamefont{{Krnjaic} et~al.}(2018)\citenamefont{{Krnjaic},
			{Machado}, and {Necib}}}]{2018PhRvD..97g5017K}
	\bibinfo{author}{\bibfnamefont{G.}~\bibnamefont{{Krnjaic}}},
	\bibinfo{author}{\bibfnamefont{P.~A.~N.} \bibnamefont{{Machado}}},
	\bibnamefont{and} \bibinfo{author}{\bibfnamefont{L.}~\bibnamefont{{Necib}}},
	\bibinfo{journal}{\prd} \textbf{\bibinfo{volume}{97}}, \bibinfo{eid}{075017}
	(\bibinfo{year}{2018}).
	
	\bibitem[{\citenamefont{{Lopes}}(2020)}]{2020ApJ...905...22L}
	\bibinfo{author}{\bibfnamefont{I.}~\bibnamefont{{Lopes}}},
	\bibinfo{journal}{\apj} \textbf{\bibinfo{volume}{905}}, \bibinfo{eid}{22}
	(\bibinfo{year}{2020}), \eprint{2101.00210}.
	
	\bibitem[{\citenamefont{{Smirnov} and {Valera}}(2021)}]{2021JHEP...09..177S}
	\bibinfo{author}{\bibfnamefont{A.~Y.} \bibnamefont{{Smirnov}}}
	\bibnamefont{and} \bibinfo{author}{\bibfnamefont{V.~B.}
		\bibnamefont{{Valera}}}, \bibinfo{journal}{Journal of High Energy Physics}
	\textbf{\bibinfo{volume}{2021}}, \bibinfo{eid}{177} (\bibinfo{year}{2021}),
	\eprint{2106.13829}.
	
	\bibitem[{\citenamefont{{Kuo} and
			{Pantaleone}}(1989{\natexlab{a}})}]{1989RvMP...61..937K}
	\bibinfo{author}{\bibfnamefont{T.~K.} \bibnamefont{{Kuo}}} \bibnamefont{and}
	\bibinfo{author}{\bibfnamefont{J.}~\bibnamefont{{Pantaleone}}},
	\bibinfo{journal}{Reviews of Modern Physics} \textbf{\bibinfo{volume}{61}},
	\bibinfo{pages}{937} (\bibinfo{year}{1989}{\natexlab{a}}).
	
	\bibitem[{\citenamefont{{Wolfenstein}}(1978)}]{1978PhRvD..17.2369W}
	\bibinfo{author}{\bibfnamefont{L.}~\bibnamefont{{Wolfenstein}}},
	\bibinfo{journal}{\prd} \textbf{\bibinfo{volume}{17}}, \bibinfo{pages}{2369}
	(\bibinfo{year}{1978}).
	
	\bibitem[{\citenamefont{{Mikheyev} and {Smirnov}}(1985)}]{1985YaFiz..42.1441M}
	\bibinfo{author}{\bibfnamefont{S.~P.} \bibnamefont{{Mikheyev}}}
	\bibnamefont{and} \bibinfo{author}{\bibfnamefont{A.~Y.}
		\bibnamefont{{Smirnov}}}, \bibinfo{journal}{Yadernaya Fizika}
	\textbf{\bibinfo{volume}{42}}, \bibinfo{pages}{1441} (\bibinfo{year}{1985}).
	
	\bibitem[{\citenamefont{{Botella} et~al.}(1987)\citenamefont{{Botella}, {Lim},
			and {Marciano}}}]{1987PhRvD..35..896B}
	\bibinfo{author}{\bibfnamefont{F.~J.} \bibnamefont{{Botella}}},
	\bibinfo{author}{\bibfnamefont{C.~S.} \bibnamefont{{Lim}}}, \bibnamefont{and}
	\bibinfo{author}{\bibfnamefont{W.~J.} \bibnamefont{{Marciano}}},
	\bibinfo{journal}{\prd} \textbf{\bibinfo{volume}{35}}, \bibinfo{pages}{896}
	(\bibinfo{year}{1987}).
	
	\bibitem[{\citenamefont{{D'olivo} et~al.}(1992)\citenamefont{{D'olivo},
			{Nieves}, and {Torres}}}]{1992PhRvD..46.1172D}
	\bibinfo{author}{\bibfnamefont{J.~C.} \bibnamefont{{D'olivo}}},
	\bibinfo{author}{\bibfnamefont{J.~F.} \bibnamefont{{Nieves}}},
	\bibnamefont{and} \bibinfo{author}{\bibfnamefont{M.}~\bibnamefont{{Torres}}},
	\bibinfo{journal}{\prd} \textbf{\bibinfo{volume}{46}}, \bibinfo{pages}{1172}
	(\bibinfo{year}{1992}).
	
	\bibitem[{\citenamefont{{Lunardini} and {Smirnov}}(2000)}]{2000NuPhB.583..260L}
	\bibinfo{author}{\bibfnamefont{C.}~\bibnamefont{{Lunardini}}} \bibnamefont{and}
	\bibinfo{author}{\bibfnamefont{A.~Y.} \bibnamefont{{Smirnov}}},
	\bibinfo{journal}{Nuclear Physics B} \textbf{\bibinfo{volume}{583}},
	\bibinfo{pages}{260} (\bibinfo{year}{2000}), \eprint{hep-ph/0002152}.
	
	\bibitem[{\citenamefont{{Konstandin} and
			{Ohlsson}}(2006)}]{2006PhLB..634..267K}
	\bibinfo{author}{\bibfnamefont{T.}~\bibnamefont{{Konstandin}}}
	\bibnamefont{and}
	\bibinfo{author}{\bibfnamefont{T.}~\bibnamefont{{Ohlsson}}},
	\bibinfo{journal}{Physics Letters B} \textbf{\bibinfo{volume}{634}},
	\bibinfo{pages}{267} (\bibinfo{year}{2006}), \eprint{hep-ph/0511010}.
	
	\bibitem[{\citenamefont{{N{\"o}tzold} and
			{Raffelt}}(1988)}]{1988NuPhB.307..924N}
	\bibinfo{author}{\bibfnamefont{D.}~\bibnamefont{{N{\"o}tzold}}}
	\bibnamefont{and}
	\bibinfo{author}{\bibfnamefont{G.}~\bibnamefont{{Raffelt}}},
	\bibinfo{journal}{Nuclear Physics B} \textbf{\bibinfo{volume}{307}},
	\bibinfo{pages}{924} (\bibinfo{year}{1988}).
	
	\bibitem[{\citenamefont{{Haxton}}(1986)}]{1986PhRvL..57.1271H}
	\bibinfo{author}{\bibfnamefont{W.~C.} \bibnamefont{{Haxton}}},
	\bibinfo{journal}{\prl} \textbf{\bibinfo{volume}{57}}, \bibinfo{pages}{1271}
	(\bibinfo{year}{1986}).
	
	\bibitem[{\citenamefont{{Parke}}(1986)}]{1986PhRvL..57.1275P}
	\bibinfo{author}{\bibfnamefont{S.~J.} \bibnamefont{{Parke}}},
	\bibinfo{journal}{\prl} \textbf{\bibinfo{volume}{57}}, \bibinfo{pages}{1275}
	(\bibinfo{year}{1986}), \eprint{2212.06978}.
	
	\bibitem[{\citenamefont{{Lopes}}(2013)}]{2013PhRvD..88d5006L}
	\bibinfo{author}{\bibfnamefont{I.}~\bibnamefont{{Lopes}}},
	\bibinfo{journal}{\prd} \textbf{\bibinfo{volume}{88}}, \bibinfo{eid}{045006}
	(\bibinfo{year}{2013}), \eprint{1308.3346}.
	
	\bibitem[{\citenamefont{{Haxton} et~al.}(2013)\citenamefont{{Haxton}, {Hamish
				Robertson}, and {Serenelli}}}]{2013ARA&A..51...21H}
	\bibinfo{author}{\bibfnamefont{W.~C.} \bibnamefont{{Haxton}}},
	\bibinfo{author}{\bibfnamefont{R.~G.} \bibnamefont{{Hamish Robertson}}},
	\bibnamefont{and} \bibinfo{author}{\bibfnamefont{A.~M.}
		\bibnamefont{{Serenelli}}}, \bibinfo{journal}{\araa}
	\textbf{\bibinfo{volume}{51}}, \bibinfo{pages}{21} (\bibinfo{year}{2013}),
	\eprint{1208.5723}.
	
	\bibitem[{\citenamefont{{Beacom} et~al.}(2017)\citenamefont{{Beacom}, {Chen},
			{Cheng}, {Doustimotlagh}, {Gao}, {Gong}, {Gong}, {Guo}, {Han}, {He}
			et~al.}}]{2017ChPhC..41b3002B}
	\bibinfo{author}{\bibfnamefont{J.~F.} \bibnamefont{{Beacom}}},
	\bibinfo{author}{\bibfnamefont{S.}~\bibnamefont{{Chen}}},
	\bibinfo{author}{\bibfnamefont{J.}~\bibnamefont{{Cheng}}},
	\bibinfo{author}{\bibfnamefont{S.~N.} \bibnamefont{{Doustimotlagh}}},
	\bibinfo{author}{\bibfnamefont{Y.}~\bibnamefont{{Gao}}},
	\bibinfo{author}{\bibfnamefont{G.}~\bibnamefont{{Gong}}},
	\bibinfo{author}{\bibfnamefont{H.}~\bibnamefont{{Gong}}},
	\bibinfo{author}{\bibfnamefont{L.}~\bibnamefont{{Guo}}},
	\bibinfo{author}{\bibfnamefont{R.}~\bibnamefont{{Han}}},
	\bibinfo{author}{\bibfnamefont{H.-J.} \bibnamefont{{He}}},
	\bibnamefont{et~al.}, \bibinfo{journal}{Chinese Physics C}
	\textbf{\bibinfo{volume}{41}}, \bibinfo{eid}{023002} (\bibinfo{year}{2017}).
	
	\bibitem[{\citenamefont{{Gonzalez-Garcia} and
			{Nir}}(2003)}]{2003RvMP...75..345G}
	\bibinfo{author}{\bibfnamefont{M.~C.} \bibnamefont{{Gonzalez-Garcia}}}
	\bibnamefont{and} \bibinfo{author}{\bibfnamefont{Y.}~\bibnamefont{{Nir}}},
	\bibinfo{journal}{Reviews of Modern Physics} \textbf{\bibinfo{volume}{75}},
	\bibinfo{pages}{345} (\bibinfo{year}{2003}), \eprint{hep-ph/0202058}.
	
	\bibitem[{\citenamefont{{Fantini} et~al.}(2018)\citenamefont{{Fantini}, {Gallo
				Rosso}, {Vissani}, and {Zema}}}]{2018arXiv180205781F}
	\bibinfo{author}{\bibfnamefont{G.}~\bibnamefont{{Fantini}}},
	\bibinfo{author}{\bibfnamefont{A.}~\bibnamefont{{Gallo Rosso}}},
	\bibinfo{author}{\bibfnamefont{F.}~\bibnamefont{{Vissani}}},
	\bibnamefont{and} \bibinfo{author}{\bibfnamefont{V.}~\bibnamefont{{Zema}}},
	\bibinfo{journal}{arXiv e-prints} \bibinfo{eid}{arXiv:1802.05781}
	(\bibinfo{year}{2018}), \eprint{1802.05781}.
	
	\bibitem[{\citenamefont{{Bahcall} and
			{Pe{\~n}a-Garay}}(2004)}]{2004NJPh....6...63B}
	\bibinfo{author}{\bibfnamefont{J.~N.} \bibnamefont{{Bahcall}}}
	\bibnamefont{and}
	\bibinfo{author}{\bibfnamefont{C.}~\bibnamefont{{Pe{\~n}a-Garay}}},
	\bibinfo{journal}{New Journal of Physics} \textbf{\bibinfo{volume}{6}},
	\bibinfo{pages}{63} (\bibinfo{year}{2004}), \eprint{hep-ph/0404061}.
	
	\bibitem[{\citenamefont{{Kumaran} et~al.}(2021)\citenamefont{{Kumaran},
			{Ludhova}, {Penek}, and {Settanta}}}]{2021Univ....7..231K}
	\bibinfo{author}{\bibfnamefont{S.}~\bibnamefont{{Kumaran}}},
	\bibinfo{author}{\bibfnamefont{L.}~\bibnamefont{{Ludhova}}},
	\bibinfo{author}{\bibfnamefont{{\"O}.}~\bibnamefont{{Penek}}},
	\bibnamefont{and}
	\bibinfo{author}{\bibfnamefont{G.}~\bibnamefont{{Settanta}}},
	\bibinfo{journal}{Universe} \textbf{\bibinfo{volume}{7}},
	\bibinfo{pages}{231} (\bibinfo{year}{2021}), \eprint{2105.13858}.
	
	\bibitem[{\citenamefont{{Tanabashi} et~al.}(2018)\citenamefont{{Tanabashi},
			{Hagiwara}, {Hikasa}, {Nakamura}, {Sumino}, {Takahashi}, {Tanaka}, {Agashe},
			{Aielli}, {Amsler} et~al.}}]{2018PhRvD..98c0001T}
	\bibinfo{author}{\bibfnamefont{M.}~\bibnamefont{{Tanabashi}}},
	\bibinfo{author}{\bibfnamefont{K.}~\bibnamefont{{Hagiwara}}},
	\bibinfo{author}{\bibfnamefont{K.}~\bibnamefont{{Hikasa}}},
	\bibinfo{author}{\bibfnamefont{K.}~\bibnamefont{{Nakamura}}},
	\bibinfo{author}{\bibfnamefont{Y.}~\bibnamefont{{Sumino}}},
	\bibinfo{author}{\bibfnamefont{F.}~\bibnamefont{{Takahashi}}},
	\bibinfo{author}{\bibfnamefont{J.}~\bibnamefont{{Tanaka}}},
	\bibinfo{author}{\bibfnamefont{K.}~\bibnamefont{{Agashe}}},
	\bibinfo{author}{\bibfnamefont{G.}~\bibnamefont{{Aielli}}},
	\bibinfo{author}{\bibfnamefont{C.}~\bibnamefont{{Amsler}}},
	\bibnamefont{et~al.}, \bibinfo{journal}{\prd} \textbf{\bibinfo{volume}{98}},
	\bibinfo{eid}{030001} (\bibinfo{year}{2018}).
	
	\bibitem[{\citenamefont{{Patrignani} et~al.}(2016)\citenamefont{{Patrignani},
			{Particle Data Group}, {Agashe}, {Aielli}, {Amsler}, {Antonelli}, {Asner},
			{Baer}, {Banerjee}, {Barnett} et~al.}}]{2016ChPhC..40j0001P}
	\bibinfo{author}{\bibfnamefont{C.}~\bibnamefont{{Patrignani}}},
	\bibinfo{author}{\bibnamefont{{Particle Data Group}}},
	\bibinfo{author}{\bibfnamefont{K.}~\bibnamefont{{Agashe}}},
	\bibinfo{author}{\bibfnamefont{G.}~\bibnamefont{{Aielli}}},
	\bibinfo{author}{\bibfnamefont{C.}~\bibnamefont{{Amsler}}},
	\bibinfo{author}{\bibfnamefont{M.}~\bibnamefont{{Antonelli}}},
	\bibinfo{author}{\bibfnamefont{D.~M.} \bibnamefont{{Asner}}},
	\bibinfo{author}{\bibfnamefont{H.}~\bibnamefont{{Baer}}},
	\bibinfo{author}{\bibfnamefont{S.}~\bibnamefont{{Banerjee}}},
	\bibinfo{author}{\bibfnamefont{R.~M.} \bibnamefont{{Barnett}}},
	\bibnamefont{et~al.}, \bibinfo{journal}{Chinese Physics C}
	\textbf{\bibinfo{volume}{40}}, \bibinfo{eid}{100001} (\bibinfo{year}{2016}).
	
	\bibitem[{\citenamefont{{Kuo} and
			{Pantaleone}}(1989{\natexlab{b}})}]{1989PhRvD..39.1930K}
	\bibinfo{author}{\bibfnamefont{T.~K.} \bibnamefont{{Kuo}}} \bibnamefont{and}
	\bibinfo{author}{\bibfnamefont{J.}~\bibnamefont{{Pantaleone}}},
	\bibinfo{journal}{\prd} \textbf{\bibinfo{volume}{39}}, \bibinfo{pages}{1930}
	(\bibinfo{year}{1989}{\natexlab{b}}).
	
	\bibitem[{\citenamefont{{Bruggen} et~al.}(1995)\citenamefont{{Bruggen},
			{Haxton}, and {Qian}}}]{1995PhRvD..51.4028B}
	\bibinfo{author}{\bibfnamefont{M.}~\bibnamefont{{Bruggen}}},
	\bibinfo{author}{\bibfnamefont{W.~C.} \bibnamefont{{Haxton}}},
	\bibnamefont{and} \bibinfo{author}{\bibfnamefont{Y.~Z.}
		\bibnamefont{{Qian}}}, \bibinfo{journal}{\prd} \textbf{\bibinfo{volume}{51}},
	\bibinfo{pages}{4028} (\bibinfo{year}{1995}).
	
	\bibitem[{\citenamefont{{de Gouv{\^e}a}}(2003)}]{2003NIMPA.503....4D}
	\bibinfo{author}{\bibfnamefont{A.}~\bibnamefont{{de Gouv{\^e}a}}},
	\bibinfo{journal}{Nuclear Instruments and Methods in Physics Research A}
	\textbf{\bibinfo{volume}{503}}, \bibinfo{pages}{4} (\bibinfo{year}{2003}),
	\eprint{hep-ph/0109150}.
	
	\bibitem[{\citenamefont{{Gouv{\^e}a} et~al.}(2000)\citenamefont{{Gouv{\^e}a},
			{Friedland}, and {Murayama}}}]{2000PhLB..490..125G}
	\bibinfo{author}{\bibfnamefont{A.~d.} \bibnamefont{{Gouv{\^e}a}}},
	\bibinfo{author}{\bibfnamefont{A.}~\bibnamefont{{Friedland}}},
	\bibnamefont{and}
	\bibinfo{author}{\bibfnamefont{H.}~\bibnamefont{{Murayama}}},
	\bibinfo{journal}{Physics Letters B} \textbf{\bibinfo{volume}{490}},
	\bibinfo{pages}{125} (\bibinfo{year}{2000}), \eprint{hep-ph/0002064}.
	
	\bibitem[{\citenamefont{{Gando} et~al.}(2011)\citenamefont{{Gando}, {Gando},
			{Ichimura}, {Ikeda}, {Inoue}, {Kibe}, {Kishimoto}, {Koga}, {Minekawa},
			{Mitsui} et~al.}}]{2011PhRvD..83e2002G}
	\bibinfo{author}{\bibfnamefont{A.}~\bibnamefont{{Gando}}},
	\bibinfo{author}{\bibfnamefont{Y.}~\bibnamefont{{Gando}}},
	\bibinfo{author}{\bibfnamefont{K.}~\bibnamefont{{Ichimura}}},
	\bibinfo{author}{\bibfnamefont{H.}~\bibnamefont{{Ikeda}}},
	\bibinfo{author}{\bibfnamefont{K.}~\bibnamefont{{Inoue}}},
	\bibinfo{author}{\bibfnamefont{Y.}~\bibnamefont{{Kibe}}},
	\bibinfo{author}{\bibfnamefont{Y.}~\bibnamefont{{Kishimoto}}},
	\bibinfo{author}{\bibfnamefont{M.}~\bibnamefont{{Koga}}},
	\bibinfo{author}{\bibfnamefont{Y.}~\bibnamefont{{Minekawa}}},
	\bibinfo{author}{\bibfnamefont{T.}~\bibnamefont{{Mitsui}}},
	\bibnamefont{et~al.}, \bibinfo{journal}{\prd} \textbf{\bibinfo{volume}{83}},
	\bibinfo{eid}{052002} (\bibinfo{year}{2011}), \eprint{1009.4771}.
	
	\bibitem[{\citenamefont{{de Holanda} et~al.}(2004)\citenamefont{{de Holanda},
			{Liao}, and {Smirnov}}}]{2004NuPhB.702..307D}
	\bibinfo{author}{\bibfnamefont{P.~C.} \bibnamefont{{de Holanda}}},
	\bibinfo{author}{\bibfnamefont{W.}~\bibnamefont{{Liao}}}, \bibnamefont{and}
	\bibinfo{author}{\bibfnamefont{A.~Y.} \bibnamefont{{Smirnov}}},
	\bibinfo{journal}{Nuclear Physics B} \textbf{\bibinfo{volume}{702}},
	\bibinfo{pages}{307} (\bibinfo{year}{2004}), \eprint{hep-ph/0404042}.
	
	\bibitem[{\citenamefont{{Casini} et~al.}(2000)\citenamefont{{Casini},
			{D'olivo}, and {Montemayor}}}]{2000PhRvD..61j5004C}
	\bibinfo{author}{\bibfnamefont{H.}~\bibnamefont{{Casini}}},
	\bibinfo{author}{\bibfnamefont{J.~C.} \bibnamefont{{D'olivo}}},
	\bibnamefont{and}
	\bibinfo{author}{\bibfnamefont{R.}~\bibnamefont{{Montemayor}}},
	\bibinfo{journal}{\prd} \textbf{\bibinfo{volume}{61}}, \bibinfo{eid}{105004}
	(\bibinfo{year}{2000}), \eprint{hep-ph/9910407}.
	
	\bibitem[{\citenamefont{{Borexino Collaboration}
			et~al.}(2018)\citenamefont{{Borexino Collaboration}, {Agostini},
			{Altenm{\"u}ller}, {Appel}, {Atroshchenko}, {Bagdasarian}, {Basilico},
			{Bellini}, {Benziger}, {Bick} et~al.}}]{2018Natur.562..505B}
	\bibinfo{author}{\bibnamefont{{Borexino Collaboration}}},
	\bibinfo{author}{\bibfnamefont{M.}~\bibnamefont{{Agostini}}},
	\bibinfo{author}{\bibfnamefont{K.}~\bibnamefont{{Altenm{\"u}ller}}},
	\bibinfo{author}{\bibfnamefont{S.}~\bibnamefont{{Appel}}},
	\bibinfo{author}{\bibfnamefont{V.}~\bibnamefont{{Atroshchenko}}},
	\bibinfo{author}{\bibfnamefont{Z.}~\bibnamefont{{Bagdasarian}}},
	\bibinfo{author}{\bibfnamefont{D.}~\bibnamefont{{Basilico}}},
	\bibinfo{author}{\bibfnamefont{G.}~\bibnamefont{{Bellini}}},
	\bibinfo{author}{\bibfnamefont{J.}~\bibnamefont{{Benziger}}},
	\bibinfo{author}{\bibfnamefont{D.}~\bibnamefont{{Bick}}},
	\bibnamefont{et~al.}, \bibinfo{journal}{\nat} \textbf{\bibinfo{volume}{562}},
	\bibinfo{pages}{505} (\bibinfo{year}{2018}).
	
	\bibitem[{\citenamefont{{Agostini} et~al.}(2019)\citenamefont{{Agostini},
			{Altenm{\"u}ller}, {Appel}, {Atroshchenko}, {Bagdasarian}, {Basilico},
			{Bellini}, {Benziger}, {Bonfini}, {Bravo} et~al.}}]{2019PhRvD.100h2004A}
	\bibinfo{author}{\bibfnamefont{M.}~\bibnamefont{{Agostini}}},
	\bibinfo{author}{\bibfnamefont{K.}~\bibnamefont{{Altenm{\"u}ller}}},
	\bibinfo{author}{\bibfnamefont{S.}~\bibnamefont{{Appel}}},
	\bibinfo{author}{\bibfnamefont{V.}~\bibnamefont{{Atroshchenko}}},
	\bibinfo{author}{\bibfnamefont{Z.}~\bibnamefont{{Bagdasarian}}},
	\bibinfo{author}{\bibfnamefont{D.}~\bibnamefont{{Basilico}}},
	\bibinfo{author}{\bibfnamefont{G.}~\bibnamefont{{Bellini}}},
	\bibinfo{author}{\bibfnamefont{J.}~\bibnamefont{{Benziger}}},
	\bibinfo{author}{\bibfnamefont{G.}~\bibnamefont{{Bonfini}}},
	\bibinfo{author}{\bibfnamefont{D.}~\bibnamefont{{Bravo}}},
	\bibnamefont{et~al.}, \bibinfo{journal}{\prd} \textbf{\bibinfo{volume}{100}},
	\bibinfo{eid}{082004} (\bibinfo{year}{2019}), \eprint{1707.09279}.
	
	\bibitem[{\citenamefont{{Bellini} et~al.}(2010)\citenamefont{{Bellini},
			{Benziger}, {Bonetti}, {Buizza Avanzini}, {Caccianiga}, {Cadonati},
			{Calaprice}, {Carraro}, {Chavarria}, {Chepurnov}
			et~al.}}]{2010PhRvD..82c3006B}
	\bibinfo{author}{\bibfnamefont{G.}~\bibnamefont{{Bellini}}},
	\bibinfo{author}{\bibfnamefont{J.}~\bibnamefont{{Benziger}}},
	\bibinfo{author}{\bibfnamefont{S.}~\bibnamefont{{Bonetti}}},
	\bibinfo{author}{\bibfnamefont{M.}~\bibnamefont{{Buizza Avanzini}}},
	\bibinfo{author}{\bibfnamefont{B.}~\bibnamefont{{Caccianiga}}},
	\bibinfo{author}{\bibfnamefont{L.}~\bibnamefont{{Cadonati}}},
	\bibinfo{author}{\bibfnamefont{F.}~\bibnamefont{{Calaprice}}},
	\bibinfo{author}{\bibfnamefont{C.}~\bibnamefont{{Carraro}}},
	\bibinfo{author}{\bibfnamefont{A.}~\bibnamefont{{Chavarria}}},
	\bibinfo{author}{\bibfnamefont{A.}~\bibnamefont{{Chepurnov}}},
	\bibnamefont{et~al.}, \bibinfo{journal}{\prd} \textbf{\bibinfo{volume}{82}},
	\bibinfo{eid}{033006} (\bibinfo{year}{2010}), \eprint{0808.2868}.
	
	\bibitem[{\citenamefont{{Abe} et~al.}(2011)\citenamefont{{Abe}, {Furuno},
			{Gando}, {Gando}, {Ichimura}, {Ikeda}, {Inoue}, {Kibe}, {Kimura}, {Kishimoto}
			et~al.}}]{2011PhRvC..84c5804A}
	\bibinfo{author}{\bibfnamefont{S.}~\bibnamefont{{Abe}}},
	\bibinfo{author}{\bibfnamefont{K.}~\bibnamefont{{Furuno}}},
	\bibinfo{author}{\bibfnamefont{A.}~\bibnamefont{{Gando}}},
	\bibinfo{author}{\bibfnamefont{Y.}~\bibnamefont{{Gando}}},
	\bibinfo{author}{\bibfnamefont{K.}~\bibnamefont{{Ichimura}}},
	\bibinfo{author}{\bibfnamefont{H.}~\bibnamefont{{Ikeda}}},
	\bibinfo{author}{\bibfnamefont{K.}~\bibnamefont{{Inoue}}},
	\bibinfo{author}{\bibfnamefont{Y.}~\bibnamefont{{Kibe}}},
	\bibinfo{author}{\bibfnamefont{W.}~\bibnamefont{{Kimura}}},
	\bibinfo{author}{\bibfnamefont{Y.}~\bibnamefont{{Kishimoto}}},
	\bibnamefont{et~al.}, \bibinfo{journal}{\prc} \textbf{\bibinfo{volume}{84}},
	\bibinfo{eid}{035804} (\bibinfo{year}{2011}), \eprint{1106.0861}.
	
	\bibitem[{\citenamefont{{Abe} et~al.}(2016)\citenamefont{{Abe}, {Haga},
			{Hayato}, {Ikeda}, {Iyogi}, {Kameda}, {Kishimoto}, {Marti}, {Miura},
			{Moriyama} et~al.}}]{2016PhRvD..94e2010A}
	\bibinfo{author}{\bibfnamefont{K.}~\bibnamefont{{Abe}}},
	\bibinfo{author}{\bibfnamefont{Y.}~\bibnamefont{{Haga}}},
	\bibinfo{author}{\bibfnamefont{Y.}~\bibnamefont{{Hayato}}},
	\bibinfo{author}{\bibfnamefont{M.}~\bibnamefont{{Ikeda}}},
	\bibinfo{author}{\bibfnamefont{K.}~\bibnamefont{{Iyogi}}},
	\bibinfo{author}{\bibfnamefont{J.}~\bibnamefont{{Kameda}}},
	\bibinfo{author}{\bibfnamefont{Y.}~\bibnamefont{{Kishimoto}}},
	\bibinfo{author}{\bibfnamefont{L.}~\bibnamefont{{Marti}}},
	\bibinfo{author}{\bibfnamefont{M.}~\bibnamefont{{Miura}}},
	\bibinfo{author}{\bibfnamefont{S.}~\bibnamefont{{Moriyama}}},
	\bibnamefont{et~al.}, \bibinfo{journal}{\prd} \textbf{\bibinfo{volume}{94}},
	\bibinfo{eid}{052010} (\bibinfo{year}{2016}).
	
	\bibitem[{\citenamefont{{Aharmim} et~al.}(2013)\citenamefont{{Aharmim},
			{Ahmed}, {Anthony}, {Barros}, {Beier}, {Bellerive}, {Beltran}, {Bergevin},
			{Biller}, {Boudjemline} et~al.}}]{2013PhRvC..88b5501A}
	\bibinfo{author}{\bibfnamefont{B.}~\bibnamefont{{Aharmim}}},
	\bibinfo{author}{\bibfnamefont{S.~N.} \bibnamefont{{Ahmed}}},
	\bibinfo{author}{\bibfnamefont{A.~E.} \bibnamefont{{Anthony}}},
	\bibinfo{author}{\bibfnamefont{N.}~\bibnamefont{{Barros}}},
	\bibinfo{author}{\bibfnamefont{E.~W.} \bibnamefont{{Beier}}},
	\bibinfo{author}{\bibfnamefont{A.}~\bibnamefont{{Bellerive}}},
	\bibinfo{author}{\bibfnamefont{B.}~\bibnamefont{{Beltran}}},
	\bibinfo{author}{\bibfnamefont{M.}~\bibnamefont{{Bergevin}}},
	\bibinfo{author}{\bibfnamefont{S.~D.} \bibnamefont{{Biller}}},
	\bibinfo{author}{\bibfnamefont{K.}~\bibnamefont{{Boudjemline}}},
	\bibnamefont{et~al.}, \bibinfo{journal}{\prc} \textbf{\bibinfo{volume}{88}},
	\bibinfo{eid}{025501} (\bibinfo{year}{2013}), \eprint{1109.0763}.
	
	\bibitem[{\citenamefont{{Cravens} et~al.}(2008)\citenamefont{{Cravens}, {Abe},
			{Iida}, {Ishihara}, {Kameda}, {Koshio}, {Minamino}, {Mitsuda}, {Miura},
			{Moriyama} et~al.}}]{2008PhRvD..78c2002C}
	\bibinfo{author}{\bibfnamefont{J.~P.} \bibnamefont{{Cravens}}},
	\bibinfo{author}{\bibfnamefont{K.}~\bibnamefont{{Abe}}},
	\bibinfo{author}{\bibfnamefont{T.}~\bibnamefont{{Iida}}},
	\bibinfo{author}{\bibfnamefont{K.}~\bibnamefont{{Ishihara}}},
	\bibinfo{author}{\bibfnamefont{J.}~\bibnamefont{{Kameda}}},
	\bibinfo{author}{\bibfnamefont{Y.}~\bibnamefont{{Koshio}}},
	\bibinfo{author}{\bibfnamefont{A.}~\bibnamefont{{Minamino}}},
	\bibinfo{author}{\bibfnamefont{C.}~\bibnamefont{{Mitsuda}}},
	\bibinfo{author}{\bibfnamefont{M.}~\bibnamefont{{Miura}}},
	\bibinfo{author}{\bibfnamefont{S.}~\bibnamefont{{Moriyama}}},
	\bibnamefont{et~al.}, \bibinfo{journal}{\prd} \textbf{\bibinfo{volume}{78}},
	\bibinfo{eid}{032002} (\bibinfo{year}{2008}), \eprint{0803.4312}.
	
	\bibitem[{\citenamefont{{Lopes}}(2017)}]{2017PhRvD..95a5023L}
	\bibinfo{author}{\bibfnamefont{I.}~\bibnamefont{{Lopes}}},
	\bibinfo{journal}{\prd} \textbf{\bibinfo{volume}{95}}, \bibinfo{eid}{015023}
	(\bibinfo{year}{2017}), \eprint{1702.00447}.
	
	\bibitem[{\citenamefont{{Berlin}}(2016)}]{2016PhRvL.117w1801B}
	\bibinfo{author}{\bibfnamefont{A.}~\bibnamefont{{Berlin}}},
	\bibinfo{journal}{\prl} \textbf{\bibinfo{volume}{117}}, \bibinfo{eid}{231801}
	(\bibinfo{year}{2016}), \eprint{1608.01307}.
	
	\bibitem[{\citenamefont{{Xing}}(2020)}]{2020PhR...854....1X}
	\bibinfo{author}{\bibfnamefont{Z.-z.} \bibnamefont{{Xing}}},
	\bibinfo{journal}{Physics Reports} \textbf{\bibinfo{volume}{854}},
	\bibinfo{pages}{1} (\bibinfo{year}{2020}), \eprint{1909.09610}.
	
	\bibitem[{\citenamefont{{B{\oe}hm} and {Fayet}}(2004)}]{2004NuPhB.683..219B}
	\bibinfo{author}{\bibfnamefont{C.}~\bibnamefont{{B{\oe}hm}}} \bibnamefont{and}
	\bibinfo{author}{\bibfnamefont{P.}~\bibnamefont{{Fayet}}},
	\bibinfo{journal}{Nuclear Physics B} \textbf{\bibinfo{volume}{683}},
	\bibinfo{pages}{219} (\bibinfo{year}{2004}), \eprint{hep-ph/0305261}.
	
	\bibitem[{\citenamefont{{Fayet}}(2007)}]{2007PhRvD..75k5017F}
	\bibinfo{author}{\bibfnamefont{P.}~\bibnamefont{{Fayet}}},
	\bibinfo{journal}{\prd} \textbf{\bibinfo{volume}{75}}, \bibinfo{eid}{115017}
	(\bibinfo{year}{2007}), \eprint{hep-ph/0702176}.
	
	\bibitem[{\citenamefont{{Ioannisian} and
			{Pokorski}}(2018)}]{2018PhLB..782..641I}
	\bibinfo{author}{\bibfnamefont{A.}~\bibnamefont{{Ioannisian}}}
	\bibnamefont{and}
	\bibinfo{author}{\bibfnamefont{S.}~\bibnamefont{{Pokorski}}},
	\bibinfo{journal}{Physics Letters B} \textbf{\bibinfo{volume}{782}},
	\bibinfo{pages}{641} (\bibinfo{year}{2018}), \eprint{1801.10488}.
	
	\bibitem[{\citenamefont{{Choi} et~al.}(2020)\citenamefont{{Choi}, {Chun}, and
			{Kim}}}]{2020PDU....3000606C}
	\bibinfo{author}{\bibfnamefont{K.-Y.} \bibnamefont{{Choi}}},
	\bibinfo{author}{\bibfnamefont{E.~J.} \bibnamefont{{Chun}}},
	\bibnamefont{and} \bibinfo{author}{\bibfnamefont{J.}~\bibnamefont{{Kim}}},
	\bibinfo{journal}{Physics of the Dark Universe}
	\textbf{\bibinfo{volume}{30}}, \bibinfo{eid}{100606} (\bibinfo{year}{2020}),
	\eprint{1909.10478}.
	
	\bibitem[{\citenamefont{{Berezhiani} and
			{Mohapatra}}(1995)}]{1995PhRvD..52.6607B}
	\bibinfo{author}{\bibfnamefont{Z.~G.} \bibnamefont{{Berezhiani}}}
	\bibnamefont{and} \bibinfo{author}{\bibfnamefont{R.~N.}
		\bibnamefont{{Mohapatra}}}, \bibinfo{journal}{\prd}
	\textbf{\bibinfo{volume}{52}}, \bibinfo{pages}{6607} (\bibinfo{year}{1995}),
	\eprint{hep-ph/9505385}.
	
	\bibitem[{\citenamefont{{Berezhiani} et~al.}(1996)\citenamefont{{Berezhiani},
			{Dolgov}, and {Mohapatra}}}]{1996PhLB..375...26B}
	\bibinfo{author}{\bibfnamefont{Z.~G.} \bibnamefont{{Berezhiani}}},
	\bibinfo{author}{\bibfnamefont{A.~D.} \bibnamefont{{Dolgov}}},
	\bibnamefont{and} \bibinfo{author}{\bibfnamefont{R.~N.}
		\bibnamefont{{Mohapatra}}}, \bibinfo{journal}{Physics Letters B}
	\textbf{\bibinfo{volume}{375}}, \bibinfo{pages}{26} (\bibinfo{year}{1996}),
	\eprint{hep-ph/9511221}.
	
	\bibitem[{\citenamefont{{Mangano} et~al.}(2006)\citenamefont{{Mangano},
			{Melchiorri}, {Serra}, {Cooray}, and {Kamionkowski}}}]{2006PhRvD..74d3517M}
	\bibinfo{author}{\bibfnamefont{G.}~\bibnamefont{{Mangano}}},
	\bibinfo{author}{\bibfnamefont{A.}~\bibnamefont{{Melchiorri}}},
	\bibinfo{author}{\bibfnamefont{P.}~\bibnamefont{{Serra}}},
	\bibinfo{author}{\bibfnamefont{A.}~\bibnamefont{{Cooray}}}, \bibnamefont{and}
	\bibinfo{author}{\bibfnamefont{M.}~\bibnamefont{{Kamionkowski}}},
	\bibinfo{journal}{\prd} \textbf{\bibinfo{volume}{74}}, \bibinfo{eid}{043517}
	(\bibinfo{year}{2006}), \eprint{astro-ph/0606190}.
	
	\bibitem[{\citenamefont{{van den Aarssen} et~al.}(2012)\citenamefont{{van den
				Aarssen}, {Bringmann}, and {Pfrommer}}}]{2012PhRvL.109w1301V}
	\bibinfo{author}{\bibfnamefont{L.~G.} \bibnamefont{{van den Aarssen}}},
	\bibinfo{author}{\bibfnamefont{T.}~\bibnamefont{{Bringmann}}},
	\bibnamefont{and}
	\bibinfo{author}{\bibfnamefont{C.}~\bibnamefont{{Pfrommer}}},
	\bibinfo{journal}{\prl} \textbf{\bibinfo{volume}{109}}, \bibinfo{eid}{231301}
	(\bibinfo{year}{2012}), \eprint{1205.5809}.
	
	\bibitem[{\citenamefont{{Adam} et~al.}(2015)\citenamefont{{Adam}, {An}, {An},
			{An}, {Anfimov}, {Antonelli}, {Baccolo}, {Baldoncini}, {Baussan}, {Bellato}
			et~al.}}]{2015arXiv150807166A}
	\bibinfo{author}{\bibfnamefont{T.}~\bibnamefont{{Adam}}},
	\bibinfo{author}{\bibfnamefont{F.}~\bibnamefont{{An}}},
	\bibinfo{author}{\bibfnamefont{G.}~\bibnamefont{{An}}},
	\bibinfo{author}{\bibfnamefont{Q.}~\bibnamefont{{An}}},
	\bibinfo{author}{\bibfnamefont{N.}~\bibnamefont{{Anfimov}}},
	\bibinfo{author}{\bibfnamefont{V.}~\bibnamefont{{Antonelli}}},
	\bibinfo{author}{\bibfnamefont{G.}~\bibnamefont{{Baccolo}}},
	\bibinfo{author}{\bibfnamefont{M.}~\bibnamefont{{Baldoncini}}},
	\bibinfo{author}{\bibfnamefont{E.}~\bibnamefont{{Baussan}}},
	\bibinfo{author}{\bibfnamefont{M.}~\bibnamefont{{Bellato}}},
	\bibnamefont{et~al.}, \bibinfo{journal}{arXiv e-prints}
	\bibinfo{eid}{arXiv:1508.07166} (\bibinfo{year}{2015}), \eprint{1508.07166}.
	
	\bibitem[{\citenamefont{{DUNE Collaboration} et~al.}(2015)\citenamefont{{DUNE
				Collaboration}, {Acciarri}, {Acero}, {Adamowski}, {Adams}, {Adamson},
			{Adhikari}, {Ahmad}, {Albright}, {Alion} et~al.}}]{2015arXiv151206148D}
	\bibinfo{author}{\bibnamefont{{DUNE Collaboration}}},
	\bibinfo{author}{\bibfnamefont{R.}~\bibnamefont{{Acciarri}}},
	\bibinfo{author}{\bibfnamefont{M.~A.} \bibnamefont{{Acero}}},
	\bibinfo{author}{\bibfnamefont{M.}~\bibnamefont{{Adamowski}}},
	\bibinfo{author}{\bibfnamefont{C.}~\bibnamefont{{Adams}}},
	\bibinfo{author}{\bibfnamefont{P.}~\bibnamefont{{Adamson}}},
	\bibinfo{author}{\bibfnamefont{S.}~\bibnamefont{{Adhikari}}},
	\bibinfo{author}{\bibfnamefont{Z.}~\bibnamefont{{Ahmad}}},
	\bibinfo{author}{\bibfnamefont{C.~H.} \bibnamefont{{Albright}}},
	\bibinfo{author}{\bibfnamefont{T.}~\bibnamefont{{Alion}}},
	\bibnamefont{et~al.}, \bibinfo{journal}{arXiv e-prints}
	\bibinfo{eid}{arXiv:1512.06148} (\bibinfo{year}{2015}), \eprint{1512.06148}.
	
	\bibitem[{\citenamefont{{An} et~al.}(2016)\citenamefont{{An}, {An}, {An},
			{Antonelli}, {Baussan}, {Beacom}, {Bezrukov}, {Blyth}, {Brugnera}, {Buizza
				Avanzini} et~al.}}]{2016JPhG...43c0401A}
	\bibinfo{author}{\bibfnamefont{F.}~\bibnamefont{{An}}},
	\bibinfo{author}{\bibfnamefont{G.}~\bibnamefont{{An}}},
	\bibinfo{author}{\bibfnamefont{Q.}~\bibnamefont{{An}}},
	\bibinfo{author}{\bibfnamefont{V.}~\bibnamefont{{Antonelli}}},
	\bibinfo{author}{\bibfnamefont{E.}~\bibnamefont{{Baussan}}},
	\bibinfo{author}{\bibfnamefont{J.}~\bibnamefont{{Beacom}}},
	\bibinfo{author}{\bibfnamefont{L.}~\bibnamefont{{Bezrukov}}},
	\bibinfo{author}{\bibfnamefont{S.}~\bibnamefont{{Blyth}}},
	\bibinfo{author}{\bibfnamefont{R.}~\bibnamefont{{Brugnera}}},
	\bibinfo{author}{\bibfnamefont{M.}~\bibnamefont{{Buizza Avanzini}}},
	\bibnamefont{et~al.}, \bibinfo{journal}{Journal of Physics G Nuclear Physics}
	\textbf{\bibinfo{volume}{43}}, \bibinfo{eid}{030401} (\bibinfo{year}{2016}),
	\eprint{1507.05613}.
	
	\bibitem[{\citenamefont{{Asplund} et~al.}(2009)\citenamefont{{Asplund},
			{Grevesse}, {Sauval}, and {Scott}}}]{2009ARA&A..47..481A}
	\bibinfo{author}{\bibfnamefont{M.}~\bibnamefont{{Asplund}}},
	\bibinfo{author}{\bibfnamefont{N.}~\bibnamefont{{Grevesse}}},
	\bibinfo{author}{\bibfnamefont{A.~J.} \bibnamefont{{Sauval}}},
	\bibnamefont{and} \bibinfo{author}{\bibfnamefont{P.}~\bibnamefont{{Scott}}},
	\bibinfo{journal}{\araa} \textbf{\bibinfo{volume}{47}}, \bibinfo{pages}{481}
	(\bibinfo{year}{2009}), \eprint{0909.0948}.
	
	\bibitem[{\citenamefont{{Capelo} and {Lopes}}(2020)}]{2020MNRAS.498.1992C}
	\bibinfo{author}{\bibfnamefont{D.}~\bibnamefont{{Capelo}}} \bibnamefont{and}
	\bibinfo{author}{\bibfnamefont{I.}~\bibnamefont{{Lopes}}},
	\bibinfo{journal}{\mnras} \textbf{\bibinfo{volume}{498}},
	\bibinfo{pages}{1992} (\bibinfo{year}{2020}), \eprint{2010.01686}.
	
	\bibitem[{\citenamefont{{Paxton} et~al.}(2019)\citenamefont{{Paxton}, {Smolec},
			{Schwab}, {Gautschy}, {Bildsten}, {Cantiello}, {Dotter}, {Farmer},
			{Goldberg}, {Jermyn} et~al.}}]{2019ApJS..243...10P}
	\bibinfo{author}{\bibfnamefont{B.}~\bibnamefont{{Paxton}}},
	\bibinfo{author}{\bibfnamefont{R.}~\bibnamefont{{Smolec}}},
	\bibinfo{author}{\bibfnamefont{J.}~\bibnamefont{{Schwab}}},
	\bibinfo{author}{\bibfnamefont{A.}~\bibnamefont{{Gautschy}}},
	\bibinfo{author}{\bibfnamefont{L.}~\bibnamefont{{Bildsten}}},
	\bibinfo{author}{\bibfnamefont{M.}~\bibnamefont{{Cantiello}}},
	\bibinfo{author}{\bibfnamefont{A.}~\bibnamefont{{Dotter}}},
	\bibinfo{author}{\bibfnamefont{R.}~\bibnamefont{{Farmer}}},
	\bibinfo{author}{\bibfnamefont{J.~A.} \bibnamefont{{Goldberg}}},
	\bibinfo{author}{\bibfnamefont{A.~S.} \bibnamefont{{Jermyn}}},
	\bibnamefont{et~al.}, \bibinfo{journal}{\apjs}
	\textbf{\bibinfo{volume}{243}}, \bibinfo{eid}{10} (\bibinfo{year}{2019}),
	\eprint{1903.01426}.
	
	\bibitem[{\citenamefont{{Turck-Chieze} and
			{Lopes}}(1993)}]{1993ApJ...408..347T}
	\bibinfo{author}{\bibfnamefont{S.}~\bibnamefont{{Turck-Chieze}}}
	\bibnamefont{and} \bibinfo{author}{\bibfnamefont{I.}~\bibnamefont{{Lopes}}},
	\bibinfo{journal}{\apj} \textbf{\bibinfo{volume}{408}}, \bibinfo{pages}{347}
	(\bibinfo{year}{1993}).
	
	\bibitem[{\citenamefont{{Bahcall} et~al.}(1995)\citenamefont{{Bahcall},
			{Pinsonneault}, and {Wasserburg}}}]{1995RvMP...67..781B}
	\bibinfo{author}{\bibfnamefont{J.~N.} \bibnamefont{{Bahcall}}},
	\bibinfo{author}{\bibfnamefont{M.~H.} \bibnamefont{{Pinsonneault}}},
	\bibnamefont{and} \bibinfo{author}{\bibfnamefont{G.~J.}
		\bibnamefont{{Wasserburg}}}, \bibinfo{journal}{Reviews of Modern Physics}
	\textbf{\bibinfo{volume}{67}}, \bibinfo{pages}{781} (\bibinfo{year}{1995}),
	\eprint{hep-ph/9505425}.
	
	\bibitem[{\citenamefont{{Bahcall} et~al.}(2006)\citenamefont{{Bahcall},
			{Serenelli}, and {Basu}}}]{2006ApJS..165..400B}
	\bibinfo{author}{\bibfnamefont{J.~N.} \bibnamefont{{Bahcall}}},
	\bibinfo{author}{\bibfnamefont{A.~M.} \bibnamefont{{Serenelli}}},
	\bibnamefont{and} \bibinfo{author}{\bibfnamefont{S.}~\bibnamefont{{Basu}}},
	\bibinfo{journal}{\apjs} \textbf{\bibinfo{volume}{165}}, \bibinfo{pages}{400}
	(\bibinfo{year}{2006}), \eprint{astro-ph/0511337}.
	
	\bibitem[{\citenamefont{{Gonzalez-Garcia}
			et~al.}(2016)\citenamefont{{Gonzalez-Garcia}, {Maltoni}, and
			{Schwetz}}}]{2016NuPhB.908..199G}
	\bibinfo{author}{\bibfnamefont{M.~C.} \bibnamefont{{Gonzalez-Garcia}}},
	\bibinfo{author}{\bibfnamefont{M.}~\bibnamefont{{Maltoni}}},
	\bibnamefont{and}
	\bibinfo{author}{\bibfnamefont{T.}~\bibnamefont{{Schwetz}}},
	\bibinfo{journal}{Nuclear Physics B} \textbf{\bibinfo{volume}{908}},
	\bibinfo{pages}{199} (\bibinfo{year}{2016}), \eprint{1512.06856}.
	
	\bibitem[{\citenamefont{{de Salas} et~al.}(2021)\citenamefont{{de Salas},
			{Forero}, {Gariazzo}, {Mart{\'\i}nez-Mirav{\'e}}, {Mena}, {Ternes},
			{T{\'o}rtola}, and {Valle}}}]{2021JHEP...02..071D}
	\bibinfo{author}{\bibfnamefont{P.~F.} \bibnamefont{{de Salas}}},
	\bibinfo{author}{\bibfnamefont{D.~V.} \bibnamefont{{Forero}}},
	\bibinfo{author}{\bibfnamefont{S.}~\bibnamefont{{Gariazzo}}},
	\bibinfo{author}{\bibfnamefont{P.}~\bibnamefont{{Mart{\'\i}nez-Mirav{\'e}}}},
	\bibinfo{author}{\bibfnamefont{O.}~\bibnamefont{{Mena}}},
	\bibinfo{author}{\bibfnamefont{C.~A.} \bibnamefont{{Ternes}}},
	\bibinfo{author}{\bibfnamefont{M.}~\bibnamefont{{T{\'o}rtola}}},
	\bibnamefont{and} \bibinfo{author}{\bibfnamefont{J.~W.~F.}
		\bibnamefont{{Valle}}}, \bibinfo{journal}{Journal of High Energy Physics}
	\textbf{\bibinfo{volume}{2021}}, \bibinfo{eid}{71} (\bibinfo{year}{2021}),
	\eprint{2006.11237}.
	
	\bibitem[{\citenamefont{{Askins} et~al.}(2020)\citenamefont{{Askins},
			{Bagdasarian}, {Barros}, {Beier}, {Blucher}, {Bonventre}, {Bourret},
			{Callaghan}, {Caravaca}, {Diwan} et~al.}}]{2020EPJC...80..416A}
	\bibinfo{author}{\bibfnamefont{M.}~\bibnamefont{{Askins}}},
	\bibinfo{author}{\bibfnamefont{Z.}~\bibnamefont{{Bagdasarian}}},
	\bibinfo{author}{\bibfnamefont{N.}~\bibnamefont{{Barros}}},
	\bibinfo{author}{\bibfnamefont{E.~W.} \bibnamefont{{Beier}}},
	\bibinfo{author}{\bibfnamefont{E.}~\bibnamefont{{Blucher}}},
	\bibinfo{author}{\bibfnamefont{R.}~\bibnamefont{{Bonventre}}},
	\bibinfo{author}{\bibfnamefont{E.}~\bibnamefont{{Bourret}}},
	\bibinfo{author}{\bibfnamefont{E.~J.} \bibnamefont{{Callaghan}}},
	\bibinfo{author}{\bibfnamefont{J.}~\bibnamefont{{Caravaca}}},
	\bibinfo{author}{\bibfnamefont{M.}~\bibnamefont{{Diwan}}},
	\bibnamefont{et~al.}, \bibinfo{journal}{European Physical Journal C}
	\textbf{\bibinfo{volume}{80}}, \bibinfo{eid}{416} (\bibinfo{year}{2020}),
	\eprint{1911.03501}.
	
	\bibitem[{\citenamefont{{Cheng} et~al.}(2017)\citenamefont{{Cheng}, {Kang},
			{Li}, {Li}, {Li}, {Yue}, {Zeng}, {Chen}, {Wu}, {Ji}
			et~al.}}]{2017ARNPS..67..231C}
	\bibinfo{author}{\bibfnamefont{J.-P.} \bibnamefont{{Cheng}}},
	\bibinfo{author}{\bibfnamefont{K.-J.} \bibnamefont{{Kang}}},
	\bibinfo{author}{\bibfnamefont{J.-M.} \bibnamefont{{Li}}},
	\bibinfo{author}{\bibfnamefont{J.}~\bibnamefont{{Li}}},
	\bibinfo{author}{\bibfnamefont{Y.-J.} \bibnamefont{{Li}}},
	\bibinfo{author}{\bibfnamefont{Q.}~\bibnamefont{{Yue}}},
	\bibinfo{author}{\bibfnamefont{Z.}~\bibnamefont{{Zeng}}},
	\bibinfo{author}{\bibfnamefont{Y.-H.} \bibnamefont{{Chen}}},
	\bibinfo{author}{\bibfnamefont{S.-Y.} \bibnamefont{{Wu}}},
	\bibinfo{author}{\bibfnamefont{X.-D.} \bibnamefont{{Ji}}},
	\bibnamefont{et~al.}, \bibinfo{journal}{Annual Review of Nuclear and Particle
		Science} \textbf{\bibinfo{volume}{67}}, \bibinfo{pages}{231}
	(\bibinfo{year}{2017}), \eprint{1801.00587}.
	
	\bibitem[{\citenamefont{{Seo}}(2019)}]{2019arXiv190305368S}
	\bibinfo{author}{\bibfnamefont{S.-H.} \bibnamefont{{Seo}}},
	\bibinfo{journal}{arXiv e-prints} \bibinfo{eid}{arXiv:1903.05368}
	(\bibinfo{year}{2019}), \eprint{1903.05368}.
	
	\bibitem[{\citenamefont{{JUNO Collaboration}}(2022)}]{2022PrPNP.12303927J}
	\bibinfo{author}{\bibnamefont{{JUNO Collaboration}}},
	\bibinfo{journal}{Progress in Particle and Nuclear Physics}
	\textbf{\bibinfo{volume}{123}}, \bibinfo{eid}{103927} (\bibinfo{year}{2022}),
	\eprint{2104.02565}.
	
	\bibitem[{\citenamefont{{SNO+ Collaboration} et~al.}(2020)\citenamefont{{SNO+
				Collaboration}, {:}, {Anderson}, {Andringa}, {Anselmo}, {Arushanova},
			{Asahi}, {Askins}, {Auty}, {Back} et~al.}}]{2020arXiv201112924S}
	\bibinfo{author}{\bibnamefont{{SNO+ Collaboration}}},
	\bibinfo{author}{\bibnamefont{{:}}}, \bibinfo{author}{\bibfnamefont{M.~R.}
		\bibnamefont{{Anderson}}},
	\bibinfo{author}{\bibfnamefont{S.}~\bibnamefont{{Andringa}}},
	\bibinfo{author}{\bibfnamefont{L.}~\bibnamefont{{Anselmo}}},
	\bibinfo{author}{\bibfnamefont{E.}~\bibnamefont{{Arushanova}}},
	\bibinfo{author}{\bibfnamefont{S.}~\bibnamefont{{Asahi}}},
	\bibinfo{author}{\bibfnamefont{M.}~\bibnamefont{{Askins}}},
	\bibinfo{author}{\bibfnamefont{D.~J.} \bibnamefont{{Auty}}},
	\bibinfo{author}{\bibfnamefont{A.~R.} \bibnamefont{{Back}}},
	\bibnamefont{et~al.}, \bibinfo{journal}{arXiv e-prints}
	\bibinfo{eid}{arXiv:2011.12924} (\bibinfo{year}{2020}), \eprint{2011.12924}.
	
	\bibitem[{\citenamefont{{Hyper-Kamiokande Proto-Collaboration}
			et~al.}(2018)\citenamefont{{Hyper-Kamiokande Proto-Collaboration}, {:},
			{Abe}, {Abe}, {Aihara}, {Aimi}, {Akutsu}, {Andreopoulos}, {Anghel}, {Anthony}
			et~al.}}]{2018arXiv180504163H}
	\bibinfo{author}{\bibnamefont{{Hyper-Kamiokande Proto-Collaboration}}},
	\bibinfo{author}{\bibnamefont{{:}}},
	\bibinfo{author}{\bibfnamefont{K.}~\bibnamefont{{Abe}}},
	\bibinfo{author}{\bibfnamefont{K.}~\bibnamefont{{Abe}}},
	\bibinfo{author}{\bibfnamefont{H.}~\bibnamefont{{Aihara}}},
	\bibinfo{author}{\bibfnamefont{A.}~\bibnamefont{{Aimi}}},
	\bibinfo{author}{\bibfnamefont{R.}~\bibnamefont{{Akutsu}}},
	\bibinfo{author}{\bibfnamefont{C.}~\bibnamefont{{Andreopoulos}}},
	\bibinfo{author}{\bibfnamefont{I.}~\bibnamefont{{Anghel}}},
	\bibinfo{author}{\bibfnamefont{L.~H.~V.} \bibnamefont{{Anthony}}},
	\bibnamefont{et~al.}, \bibinfo{journal}{arXiv e-prints}
	\bibinfo{eid}{arXiv:1805.04163} (\bibinfo{year}{2018}), \eprint{1805.04163}.
	
	\bibitem[{\citenamefont{{Acciarri} et~al.}(2016)\citenamefont{{Acciarri},
			{Acero}, {Adamowski}, {Adams}, {Adamson}, {Adhikari}, {Ahmad}, {Albright},
			{Alion}, {Amador} et~al.}}]{2016arXiv160102984A}
	\bibinfo{author}{\bibfnamefont{R.}~\bibnamefont{{Acciarri}}},
	\bibinfo{author}{\bibfnamefont{M.~A.} \bibnamefont{{Acero}}},
	\bibinfo{author}{\bibfnamefont{M.}~\bibnamefont{{Adamowski}}},
	\bibinfo{author}{\bibfnamefont{C.}~\bibnamefont{{Adams}}},
	\bibinfo{author}{\bibfnamefont{P.}~\bibnamefont{{Adamson}}},
	\bibinfo{author}{\bibfnamefont{S.}~\bibnamefont{{Adhikari}}},
	\bibinfo{author}{\bibfnamefont{Z.}~\bibnamefont{{Ahmad}}},
	\bibinfo{author}{\bibfnamefont{C.~H.} \bibnamefont{{Albright}}},
	\bibinfo{author}{\bibfnamefont{T.}~\bibnamefont{{Alion}}},
	\bibinfo{author}{\bibfnamefont{E.}~\bibnamefont{{Amador}}},
	\bibnamefont{et~al.}, \bibinfo{journal}{arXiv e-prints}
	\bibinfo{eid}{arXiv:1601.02984} (\bibinfo{year}{2016}), \eprint{1601.02984}.
	
\end{thebibliography}

\end{document}